\documentclass[aps,prb,reprint,superscriptaddress,floatfix]{revtex4-1}

\usepackage[pdftex]{graphicx}
\usepackage{color,soul}
\usepackage{enumitem}
\usepackage{gensymb}
\usepackage{epstopdf}
\usepackage{bm}
\usepackage{amsmath,amsthm,amssymb}
\usepackage{dcolumn}
\usepackage{xr}

\begin{document}

\title{Skyrmions and spirals in MnSi under hydrostatic pressure}
\author{L.J. Bannenberg}
\affiliation{Faculty of Applied Sciences, Delft University of Technology, Mekelweg 15, 2629 JB Delft, The Netherlands}
\email{l.j.bannenberg@tudelft.nl}
\author{R. Sadykov}
\affiliation{Institute for Nuclear Research, Russian Academy of Sciences, 60-Letiya Oktiabria 7a, 117312 Moscow, Russia}
\affiliation{Institute of High Pressure Physics, Russian Academy of Sciences, 142190 Troitsk, Russia}
\affiliation{National University of Science and Technology MISiS, Leninsky prosp. 4, Moscow, 119049, Russia}
\author{R.M. Dalgliesh}
\affiliation{ISIS neutron source, Rutherford Appleton Laboratory, STFC, OX11 0QX Didcot, United Kingdom}
\author{C. Goodway}
\affiliation{ISIS neutron source, Rutherford Appleton Laboratory, STFC, OX11 0QX Didcot, United Kingdom}
\author{D.L. Schlagel}
\affiliation{Ames Laboratory, Iowa State University, Ames, IA 50011, USA}
\author{T.A. Lograsso}
\affiliation{Ames Laboratory, Iowa State University, Ames, IA 50011, USA}
\affiliation{Department of Materials Science and Engineering, Iowa State University, Ames IA 50011 USA}
\author{P. Falus}
\affiliation{Institut Laue-Langevin, 71 Avenue des Martyrs, CS 20156 Grenoble, France}
\author{E. Leli\`{e}vre-Berna}
\affiliation{Institut Laue-Langevin, 71 Avenue des Martyrs, CS 20156 Grenoble, France}
\author{A. O. Leonov}
\affiliation{Chiral Research Center, Hiroshima University, Higashi Hiroshima, Hiroshima 739-8526, Japan}
\author{C. Pappas}
\affiliation{Faculty of Applied Sciences, Delft University of Technology, Mekelweg 15, 2629 JB Delft, The Netherlands}

\date{\today}

\begin{abstract}
The archetype cubic chiral magnet MnSi is home to some of the most fascinating states in condensed matter such as skyrmions and a non-Fermi liquid behavior in conjunction with a topological Hall effect under hydrostatic pressure. Using small angle neutron scattering, we study the evolution of the helimagnetic, conical and skyrmionic correlations with increasing hydrostatic pressure. We show that the helical propagation vector smoothly reorients from $\langle 111 \rangle$ to $\langle100\rangle$ at intermediate pressures. At higher pressures, above the critical pressure, the long-range helimagnetic order disappears at zero magnetic field. Nevertheless, skyrmion lattices and conical spirals form under magnetic fields, in a part of the phase diagram where a topological Hall effect and a non-Fermi liquid  behavior have been reported.  These unexpected results shed light on the puzzling behavior of MnSi at high pressures and the mechanisms that destabilize the helimagnetic long-range order at the critical pressure.
\end{abstract}

\maketitle

\section{Introduction}

A well-known route to discover new and exotic states of matter is by tweaking the magnetic interactions through chemical substitution or hydrostatic pressure. A prominent example is the archetype chiral cubic magnet MnSi that hosts some of the most peculiar states reported in condensed matter physics, such as skyrmion lattices \cite{muhlbauer2009,neubauer2009} and under hydrostatic pressure a non-Fermi liquid (NFL) phase \cite{pfleiderer2001,doiron2003}. The NFL phase is characterized by a $T^{3/2}$  temperature dependence of the resistivity and emerges for $p$ $>$ $p_C$ $\approx$ 1.4~GPa without quantum criticality \cite{pfleiderer2007}.  In addition, this phase is home to a sizable topological Hall effect (THE) \cite{lee2009,ritz2013,ritz2013prb}, a key characteristic of topological non-trivial magnetic order. For $p$ $>$ $p_C$ the long-range helimagnetic order at zero field is suppressed and a partial helimagnetic order has been reported \cite{pfleiderer2004,pintschovius2004}, the nature of which and its relation to the NFL and THE signals remains an open question. 

\begin{figure}[tb]
\begin{center}
\includegraphics[width= 0.42\textwidth]{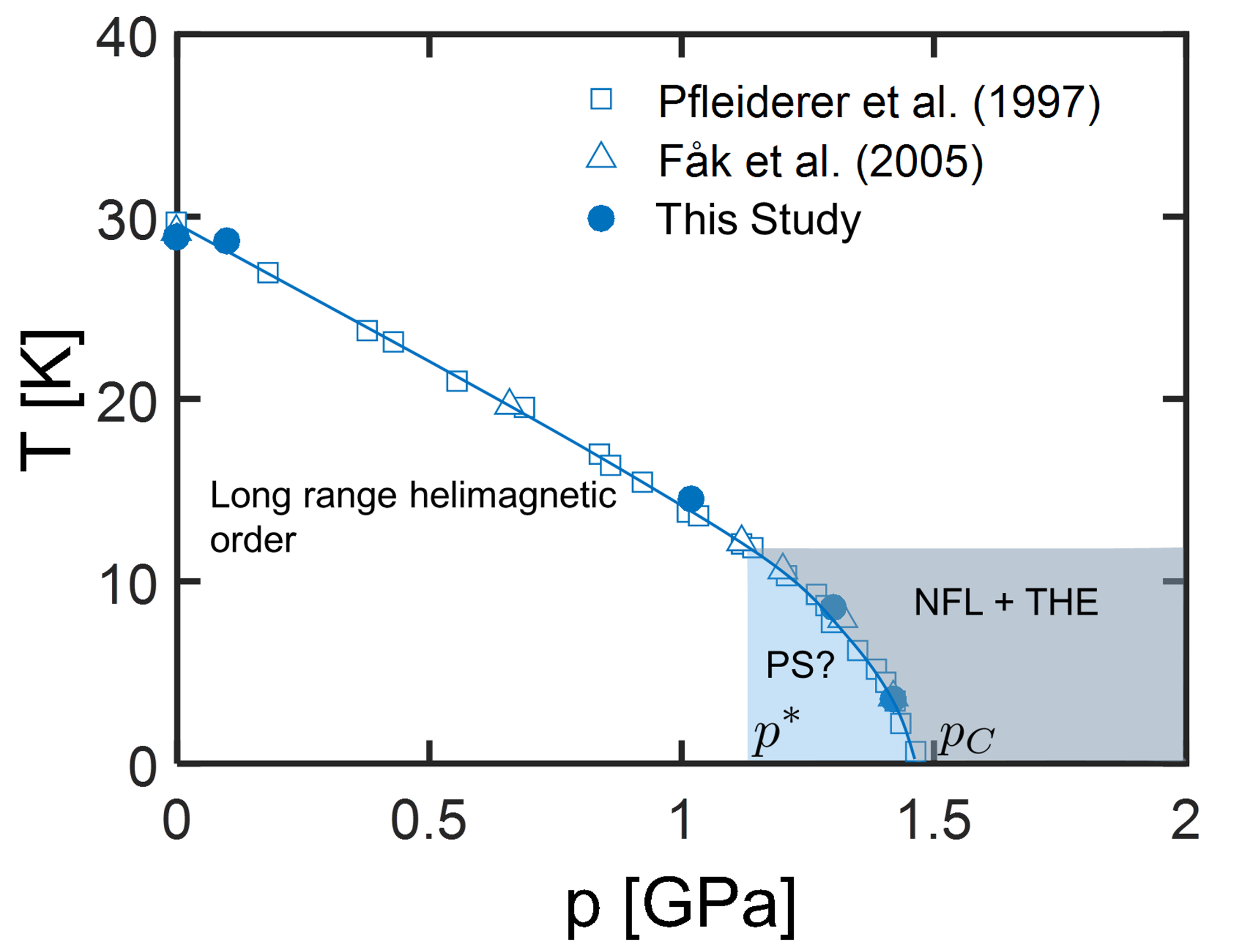}
\caption{Zero magnetic field phase diagram of MnSi under hydrostatic pressure. The helimagnetic transition temperature is after Refs. \cite{pfleiderer1997} and \cite{fak2005}. 
The non-Fermi liquid (NFL) and topological Hall effect (THE) behavior appears for pressures higher than $p^*$ and $p_C$ is the critical pressure, above which the helimagnetic order disappears. $p^*$ also marks the onset of the alleged phase separated state (PS) where helimagnetic and paramagnetic volumes may coexist. }
\label{Phase_Diagram}
\end{center}
\end{figure}

\begin{figure*}[tb]
\begin{center}
\includegraphics[width= 1\textwidth]{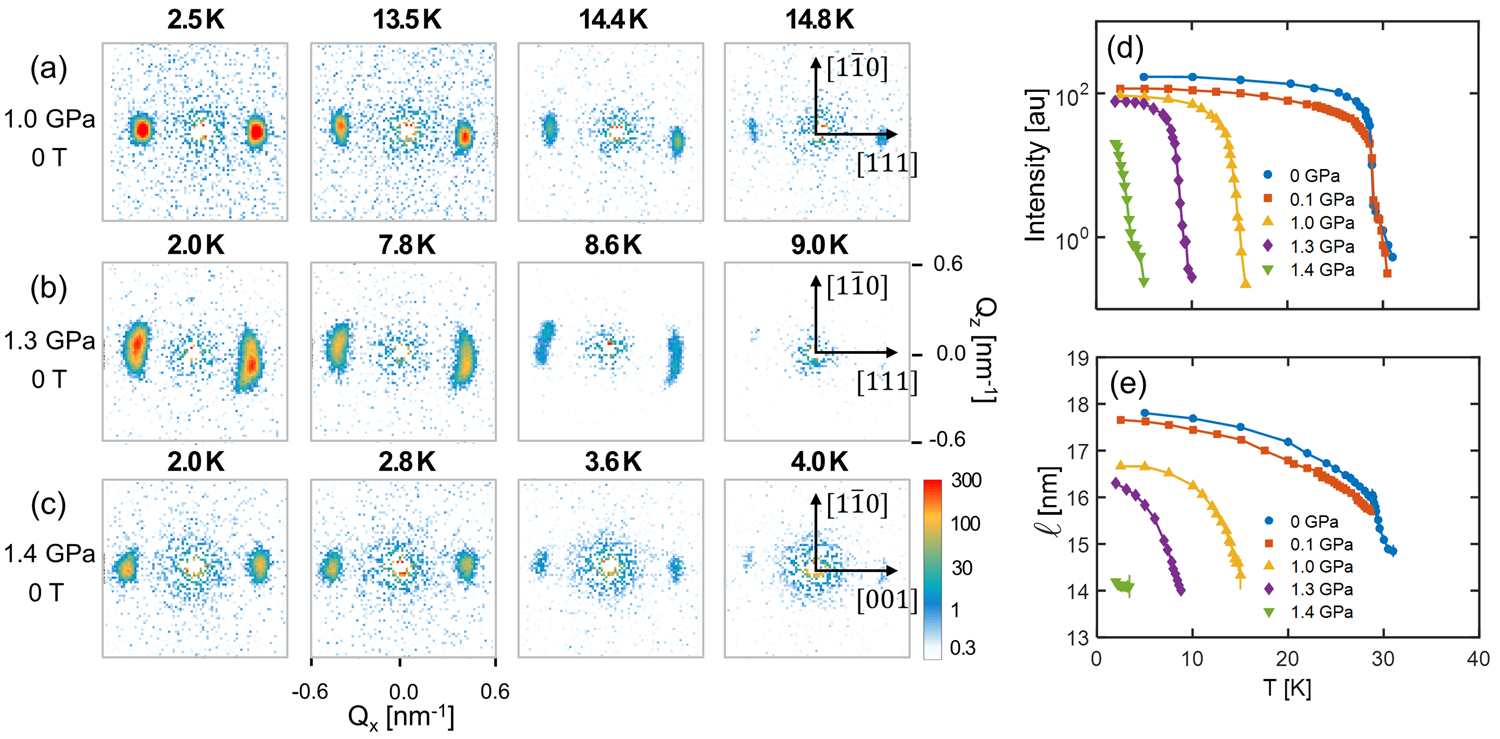}
\caption{SANS results at zero magnetic field. Panels (a) - (c) display characteristic SANS patterns obtained at a pressure of (a) $p$ = 1.0~GPa, (b) 1.3~GPa and (c) 1.4~GPa and for the temperatures indicated. Panel (d) displays the total scattered intensity as a function of temperature for the pressures indicated in the figure legend. The total scattered intensity is obtained by summing all the intensity of the SANS patterns. Panel (e) shows the temperature dependence of the pitch of the helix $\ell$ for the pressures indicated in the figure legend. $\ell$ is obtained from the maximum of a fit of a Gaussian to the radial averaged SANS patterns.}
\label{ZeroField}
\end{center}
\end{figure*}

Substantial qualitative changes to the ground state of MnSi already occur at pressures lower than $p_C$. The NFL and THE behavior for $T$ $>$ $T_C$ has been reported for $p$ $>$ $p^*$ $\approx$ 1.2~GPa, as illustrated by Fig. \ref{Phase_Diagram}. Furthermore, there is controversy about whether the region of the phase diagram between $p^*$ and $p_C$ is characterized by phase separation between helimagnetic and paramagnetic volumes, \cite{yu2004,uemura2007}, or not \cite{andreica2010}. 

In this paper, we revisit the magnetic phase diagram of MnSi under hydrostatic pressure and elucidate the evolution of the helimagnetic correlations around $p^*$ and $p_C$. For this reason, we performed small angle neutron scattering (SANS) experiments as a function of pressure and systematically applied the magnetic field both parallel and perpendicular to the incoming neutron beam, designated by its wavevector $\vec{k}_i$. In this way, we obtain a complete overview of the helimagnetic, conical and skyrmionic correlations that has not been provided by previous studies. 

We find that at zero magnetic field the helical propagation vector $\vec{\tau}$ reorients around $p^*$ from $\langle111\rangle$  at low pressures to $\langle100\rangle$ at high pressures. Furthermore, although the long-range helimagnetic order disappears for $p$ $>$ $p_C$, a magnetic field induces long-range skyrmion lattices and conical spirals even for $p$ $>$ $p_C$, in a region of the phase diagram governed by NFL and THE behavior. These unexpected results provide a strong indication that at high pressures the magnetic moment depends strongly on the strength of the magnetic field. Hydrostatic pressure would therefore soften the magnetic moment and consequently  enhance the itinerant electron character of the magnetism of MnSi, an effect that could play a role in the destabilization of the helimagnetic long-range order at $p_C$.

\section{Experimental}
The SANS measurements were performed at the time-of-flight instrument Larmor of the ISIS Neutron Source and on a 110~mg single crystal of MnSi. These measurements were complemented by high resolution neutron spin echo (NSE) spectroscopy measurements at the IN15 spectrometer of the Institut Laue-Langevin. In both experiments, the sample was aligned with the [1$\bar{1}$0] crystallographic direction vertical and positioned in the clamp type pressure cell described in Ref. \cite{sadykov2018}. More experimental details are provided in the Supplemental Material \cite{SI}.

\begin{figure*}[tb]
\begin{center}
\includegraphics[width= 1\textwidth]{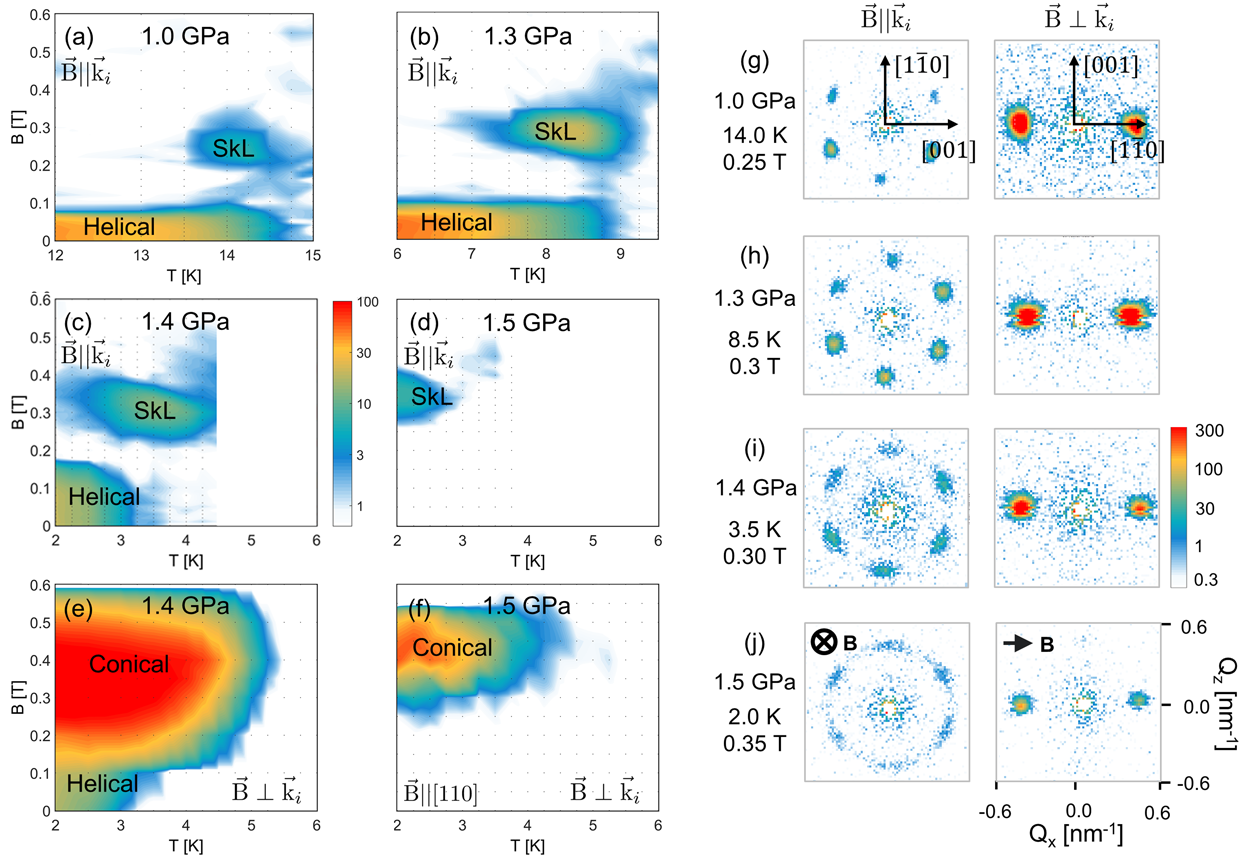}
\caption{SANS results under magnetic field. Panels (a) - (f) display contour plots of the total scattered SANS intensity, as obtained from summing the intensity of the SANS patterns outside the direct beam, as a function of temperature and magnetic field and for the different pressures indicated. The magnetic field was in Panels (a) - (d) applied parallel to the incoming neutron beam designated by $\vec{k}_i$ ($\vec{B} || \vec{k_i}$) and in Panels (e) and (f) it was applied perpendicular to it ($\vec{B} \perp \vec{k_i}$). The gray dots indicate the points at which a measurement was performed. Panels (g) - (j) display characteristic SANS patterns for the indicated pressures, temperatures and magnetic fields and for the two experimental configurations. The measurements were performed after zero field cooling the sample and by stepwise increasing the magnetic field.}
\label{MagneticField}
\end{center}
\end{figure*}

\section{Results and discussion} 
\subsection{Zero Field}
Figure \ref{ZeroField}(a)-(c) displays characteristic SANS patterns collected at zero magnetic field and for different temperatures and pressures. The patterns for $p$ = 1.0~GPa are very similar to the ones seen at zero pressure: below $T_C$ $\approx$ 14~K long-range helical order sets-in with $\vec{\tau}$ $\parallel$  $\langle111\rangle$. However, above $T_C$, we do not observe the intense ring of scattering seen at ambient pressures, and which is characteristic of the  isotropic, chiral and short-range helimagnetic correlations seen in the precursor phase \cite{grigoriev2005,pappas2009,janoschek2013,pappas2017,bannenberg2017}. In addition, the temperature evolution of the total scattered intensity displayed in Fig. \ref{ZeroField}(d), obtained by summing the intensity over the entire SANS pattern, is different from that at lower pressures. For $p$ = 0 and 0.1~GPa there is a clear kink at $T_C$, which is completely absent at higher pressures. Therefore, we conclude that this diffuse scattering weakens considerably with increasing pressure and at high pressures it does not exceed the background of the pressure cell. This substantial weakening of the diffuse scattering indicates a suppression of the precursor phase. This conclusion is supported by the NSE spectra presented in the supplement \cite{SI} (Fig. \ref{NSE}), which under pressure do not show the fluctuations associated with the precursor phase but remain elastic on the nanoseconds timescale even for  $T \gtrsim T_C$. This result is also consistent with the disappearance under pressure of the `shoulder', characteristic of the precursor phase, observed slightly above $T_C$  in the temperature dependence of the resistivity and specific heat  \cite{petrova2012, sidorov2014}.

When the pressure reaches 1.3~GPa, the helical Bragg peaks broaden substantially [Fig. \ref{ZeroField}(b)], implying that the direction of $\vec{\tau}$ is relatively ill-defined. Surprisingly, if the pressure is further increased to 1.4~GPa, the  peaks become again as sharp as for lower pressures. However, the helices are no longer aligned along $\langle111\rangle$ but along $\langle100\rangle$ (see Fig. \ref{ZeroField}(c) and the rocking scans of Fig. \ref{RockingCurve}). This unexpected and novel result reveals that the direction of $\vec{\tau}$ crosses over around $p^*$. This effect can be accounted for by the cubic anisotropy, $f_{a1} = K(m_x^4+m_y^4+m_z^4)$, which has only minima at $\vec{\tau}$ $||$ $\langle 111 \rangle$ for $K$ $<$ 0 and $\vec{\tau}$ $||$ $\langle 100 \rangle$ for $K$ $>$ 0 \cite{bak1980}. Our results thus reveal that $K$ changes sign around $p^*$, the region of the phase diagram where Larmor diffraction indicates an abrupt change of the spontaneous magnetostriction \cite{pfleiderer2007}. 

The change of the sign of $K$ is accompanied by  a relative weakening of the ferromagnetic exchange with respect to the Dzyaloshinsky-Moriya interaction. This is revealed by the decrease of the pitch of the helix $\ell$ $\propto$ $J/D$, shown in Fig. \ref{ZeroField}(e), which is consistent with earlier measurements \cite{fak2005}. In addition, we note that the total scattered intensity at low temperatures does not reduce dramatically with increasing pressure, not even for $p$ $>$ $p^*$ [Fig. \ref{ZeroField}(d)]. Therefore, the magnetic moment does not vanish, which is in agreement with previous studies \cite{thessieu1997,pfleiderer1997,petrova2006,otero2009}.  Finally, the temperature dependence of the total scattered intensity also indicates that the weak first-order phase transition, as seen at ambient pressure \cite{stishov2007,pappas2009,janoschek2013,pappas2017}, persists up to $p_C$.

\begin{figure}[tb]
\begin{center}
\includegraphics[width= 0.42\textwidth]{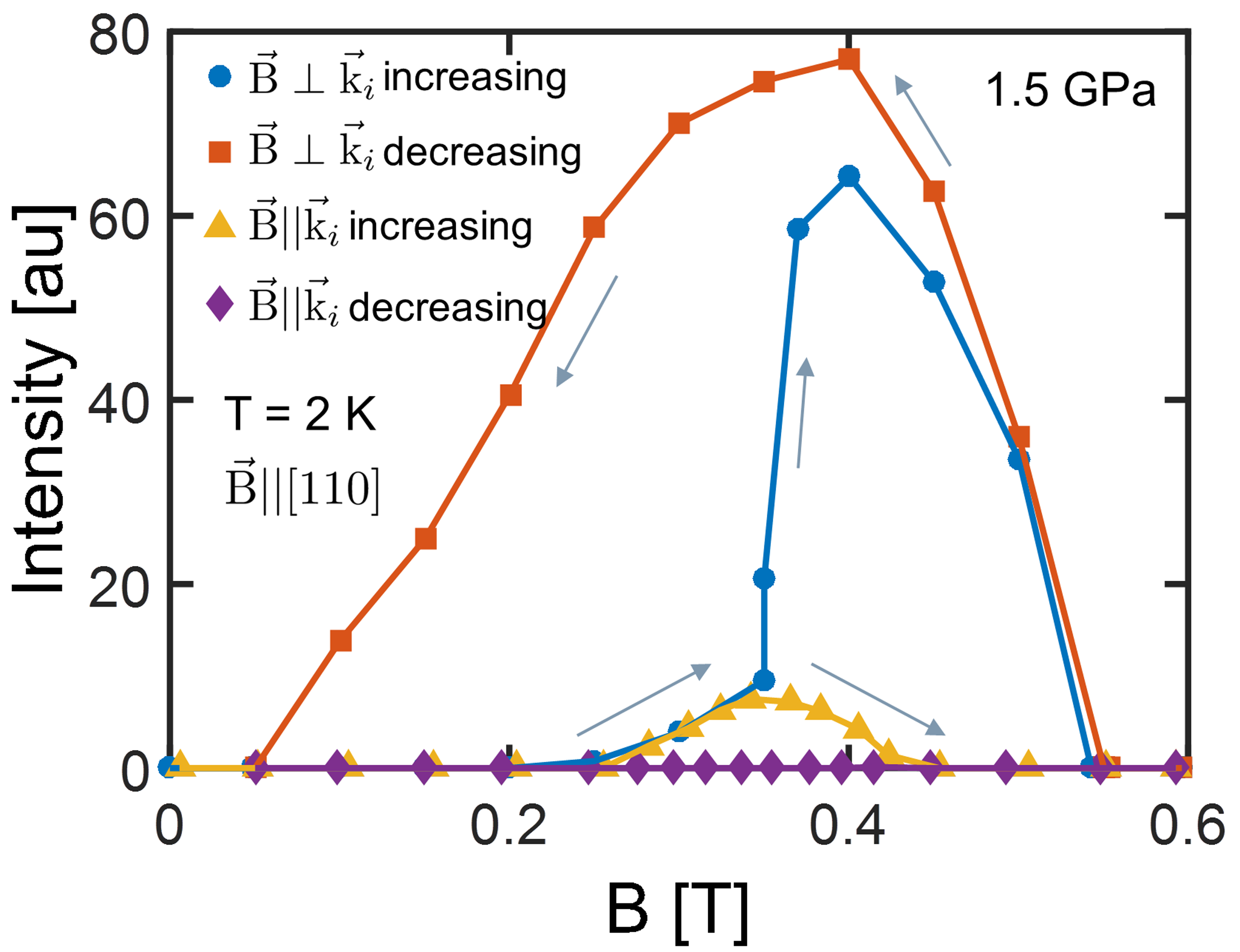}
\caption{Magnetic field dependence of the total scattered intensity, as obtained from summing the intensity of the SANS patterns outside the direct beam, at $T$ = 2~K and $p$ = 1.5~GPa. The magnetic field was applied along the $[1\bar{1}0]$ crystallographic direction and both parallel ($\vec{B} || \vec{k_i}$) and perpendicular ($\vec{B} \perp \vec{k_i}$) to the incoming neutron beam. The measurements were performed after zero field cooling the sample and as a function of both increasing and decreasing magnetic field.}
\label{1p5GPa_History}
\end{center}
\end{figure}

\subsection{Magnetic Field}
The results under magnetic field are summarized by Fig. \ref{MagneticField}, which displays SANS patterns and contour plots of the total scattered intensity as a function of temperature and magnetic field at different pressures and for both experimental configurations: $\vec{B} || \vec{k_i}$  and $\vec{B} \perp \vec{k_i}$. These results clearly show three distinct regimes: $p$ $<$ $p^*$, $p^*$ $<$ $p$ $<$ $p_C$ and $p$ $>$ $p_C$.

For $p$ $<$ $p^*$, the magnetic field-temperature phase diagrams remain qualitatively the same as at zero pressure. This is illustrated by the contour plot of Fig. \ref{MagneticField}(a), which displays the total scattered intensity for $p$ = 1.0~GPa and for $\vec{B} || \vec{k_i}$, the configuration sensitive to helical modulations perpendicular to the magnetic field. This plot shows two regions with intensity: at low magnetic fields and below $T_C$ owing to the helical phase, and at intermediate magnetic fields, in the A-phase region, due to the six-fold symmetric SANS patterns [Fig. \ref{MagneticField}(g)] characteristic of the skyrmion lattice phase \cite{muhlbauer2009}. In the complementary configuration with $\vec{B} \perp \vec{k_i}$, which is sensitive to helimagnetic correlations parallel to the magnetic field, the SANS patterns show two peaks along the field [Fig. \ref{MagneticField}(g)], characteristic of the conical phase, which coexists with the skyrmion lattice in the A-phase.

The behavior changes considerably above $p^*$. For $p^*$ $<$ $p$ $<$ $p_C$  the skyrmion lattice appears at temperatures significantly higher than $T_C$ [Fig. \ref{MagneticField}(b),(c)]. At $p$ = 1.3~GPa $>$ $p^*$, the skyrmion lattice phase persists up to $T$ = 9.4 ~K, i.e. $T_C$ + 0.4~K. At a slightly larger pressure of 1.4~GPa, the skyrmion lattice extends up to $T$ = 4.5, i.e. $T_C$ + 1.5~K, which is a most remarkable result. A similar but even more pronounced effect is seen for the conical phase, which as shown by the contour plot of the total scattered intensity for $\vec{B} \perp \vec{k_i}$ [Fig. \ref{MagneticField}(e)], persists up to $T$ $\approx$ 5.5~K, i.e. up to $\approx$ 2$T_C$.

Above $p_C$, the skyrmion lattice and conical spiral scattering appears under magnetic fields despite the absence of any scattering at zero  field. This is illustrated by the patterns of Fig. \ref{Patterns_2K_1p5Gpa} and the contour plots of Fig. \ref{MagneticField}(d),(f) for $p$ = 1.5~GPa, which show that the skyrmion lattice scattering appears up to $T$ = 3~K [Fig. \ref{MagneticField}(d)], whereas the conical scattering persists up to temperatures as high as $T$ = 4.5~K [Fig. \ref{MagneticField}(f)]. The characteristic skyrmion lattice pattern of six peaks is seen for $\vec{B}$ $||$ $\vec{k}_i$ [Fig. \ref{MagneticField}(j)],  superimposed on a weak ring of scattering. These six maxima are noticeably broader along the ring than at lower pressures [Fig. \ref{MagneticField}(g)], suggesting that the sixth-order anisotropy, which couples  the skyrmion to the crystallographic lattice \cite{muhlbauer2009,bannenberg2017reorientations},  weakens considerably at these high pressures.  On the other hand, the width of both the peaks and the ring is limited by the experimental resolution. Among the six skyrmion lattice maxima, two are oriented along the $\langle110\rangle$-crystallographic directions and their combination with the ring of scattering thus closely resembles the scattering that was previously attributed to partial helimagnetic order \cite{pintschovius2004,pfleiderer2004}.

The intensity and extent of both the skyrmion lattice and conical scattering strongly depend on the magnetic history. As shown in Fig. \ref{1p5GPa_History}, when the field is increased, skyrmion scattering appears between 0.22 and 0.43~T whereas it does not reappear when the magnetic field is decreased from 0.8~T, i.e. from the field-polarized state. On the other hand, with increasing field, the conical scattering appears at the same magnetic field as the skyrmion lattice phase, increases rapidly at $\sim$0.4~T before vanishing at the  field-polarized state. However, when the magnetic field is decreased from 0.8~T to zero, the conical scattering reappears, becomes even more intense, in particular below $\sim$0.4~T, and extends down to very low fields. This strong hysteric behavior suggests that sizable energy barriers are involved in the nucleation of these field-induced phases.

The formation under magnetic field of skyrmion lattices and conical spirals at $T>T_C$ for $p^*$ $<$ $p$ $<$ $p_C$ and at $T>0$ for $p$ $>$ $p_C$ is a remarkable observation, which sheds light on earlier puzzling magnetization results \cite{thessieu1997}. In the absence of  long-range helimagnetic correlations at zero field, an important question is that of the ground state out of which a magnetic field can stabilize skyrmion lattices and conical spirals. If this state would consist of a partial helimagnetic order, as previously suggested \cite{pfleiderer2004,pintschovius2004}, one would expect a considerable amount of scattering to persist down to zero magnetic field especially when the magnetic field is reduced from high to low values (Fig. \ref{1p5GPa_History}). However, this is not what we observe. One may argue that we are hindered by the background of the pressure cell that would mask the scattering of the partial helical order. On the other hand, as seen in Fig. \ref{ZeroField}(d), the background of the cell is small enough to let us observe the diffuse scattering of the precursor phase at low pressures, which is more than two orders of magnitude weaker than the maximum of the conical scattering at 1.5~GPa (Fig. \ref{1p5GPa_History}).  We thus conclude that we should have been able to observe the partial helical order if the intensity of the scattering would have been comparable to that of the diffuse scattering of the precursor phase at low pressures. 
 
The destabilization of the  long-range helical order at $p_C$ and the nature of the phase that sets-in above $p_C$ have been widely debated in the literature \cite{rossler2006, BlueQuantumFog2006, Binz2006, Fischer2008, Hopkinson2009, Kirkpatrick2010, Pfleiderer2009dr}. In the absence of quantum criticality \cite{pfleiderer2007} it has been suggested that the  NFL and THE behavior  would be the signature of spin textures and spin excitations with a non-trivial topology \cite{Pfleiderer2009dr}. These would predominately be stabilized  by magnetic moment modulus fluctuations, resulting from the reduced longitudinal spin stiffness and the weak itinerant electron character of the magnetism in MnSi \cite{rossler2006}. The softening of the magnetic moment  implies that the magnetization modulus  becomes strongly field dependent, and this could explain the formation of  conical spirals or skyrmion lattices under magnetic field, even in the absence of any long-range helimagnetic correlations at zero field \cite{leonov2018b}.  Following this scenario, hydrostatic pressure would reduce the longitudinal spin stiffness and soften the magnetic moment. In other words, hydrostatic pressure would enhance the itinerant electron character of the magnetism in MnSi and this enhancement could be a driver in the destabilization of the long-range helical order at $p_C$. 

\section{Conclusion} 
In conclusion, our SANS results shed light on the puzzling behavior of MnSi under pressure. With increasing pressure, first the helical propagation vector smoothly reorients from $\langle 111 \rangle$ to $\langle100\rangle$ around $p^*$. At higher pressures, the helimagnetic order persists at zero magnetic field up to $p_C$.  Nevertheless, magnetic fields stabilize conical spirals and skyrmion lattices even in the absence of helimagnetic correlations at zero field, thus even above $p_C$ in a part of the phase diagram  where previous studies reported NFL and THE behavior. This unexpected observation can possibly be attributed to a softening of the magnetic moment reflecting an enhancement of the itinerant electron character of the magnetism in MnSi with increasing pressure. We argue that this enhancement could be a driver in the destabilization of the long-range helical order at $p_C$.

\nocite{petrova2012,sidorov2014,pappas2017,bannenberg2017reorientations,bannenberg2018mnfesisquid,bannenberg2018mnfesisans,sadykov2018}
				
\begin{acknowledgments}
The authors wish to thank N. Martin for fruitful discussions and the ISIS and ILL support staff for their assistance. Experiments at the ISIS Pulsed Neutron and Muon Source were supported by a beamtime allocation from the Science and Technology Facilities Council and The Netherlands Organization for Scientific Research (NWO). The ISIS experimental data is available through doi.org/10.5286/ISIS.E.90588288. The work of LJB and CP is financially supported by The Netherlands Organization for Scientific Research through project 721.012.102 (Larmor). TAL and DLS acknowledge support by the U.S. Department of Energy, Basic Energy Sciences under Contract  DE-AC02-07CH11358. 
\end{acknowledgments}

\bibliography{MnSi_Cu2OSeO3_Rotation}

\begin{thebibliography}{40}%
\makeatletter
\providecommand \@ifxundefined [1]{%
 \@ifx{#1\undefined}
}%
\providecommand \@ifnum [1]{%
 \ifnum #1\expandafter \@firstoftwo
 \else \expandafter \@secondoftwo
 \fi
}%
\providecommand \@ifx [1]{%
 \ifx #1\expandafter \@firstoftwo
 \else \expandafter \@secondoftwo
 \fi
}%
\providecommand \natexlab [1]{#1}%
\providecommand \enquote  [1]{``#1''}%
\providecommand \bibnamefont  [1]{#1}%
\providecommand \bibfnamefont [1]{#1}%
\providecommand \citenamefont [1]{#1}%
\providecommand \href@noop [0]{\@secondoftwo}%
\providecommand \href [0]{\begingroup \@sanitize@url \@href}%
\providecommand \@href[1]{\@@startlink{#1}\@@href}%
\providecommand \@@href[1]{\endgroup#1\@@endlink}%
\providecommand \@sanitize@url [0]{\catcode `\\12\catcode `\$12\catcode
  `\&12\catcode `\#12\catcode `\^12\catcode `\_12\catcode `\%12\relax}%
\providecommand \@@startlink[1]{}%
\providecommand \@@endlink[0]{}%
\providecommand \url  [0]{\begingroup\@sanitize@url \@url }%
\providecommand \@url [1]{\endgroup\@href {#1}{\urlprefix }}%
\providecommand \urlprefix  [0]{URL }%
\providecommand \Eprint [0]{\href }%
\providecommand \doibase [0]{http://dx.doi.org/}%
\providecommand \selectlanguage [0]{\@gobble}%
\providecommand \bibinfo  [0]{\@secondoftwo}%
\providecommand \bibfield  [0]{\@secondoftwo}%
\providecommand \translation [1]{[#1]}%
\providecommand \BibitemOpen [0]{}%
\providecommand \bibitemStop [0]{}%
\providecommand \bibitemNoStop [0]{.\EOS\space}%
\providecommand \EOS [0]{\spacefactor3000\relax}%
\providecommand \BibitemShut  [1]{\csname bibitem#1\endcsname}%
\let\auto@bib@innerbib\@empty
\bibitem [{\citenamefont {M{\"u}hlbauer}\ \emph {et~al.}(2009)\citenamefont
  {M{\"u}hlbauer}, \citenamefont {Binz}, \citenamefont {Jonietz}, \citenamefont
  {Pfleiderer}, \citenamefont {Rosch}, \citenamefont {Neubauer}, \citenamefont
  {Georgii},\ and\ \citenamefont {B{\"o}ni}}]{muhlbauer2009}%
  \BibitemOpen
  \bibfield  {author} {\bibinfo {author} {\bibfnamefont {S.}~\bibnamefont
  {M{\"u}hlbauer}}, \bibinfo {author} {\bibfnamefont {B.}~\bibnamefont {Binz}},
  \bibinfo {author} {\bibfnamefont {F.}~\bibnamefont {Jonietz}}, \bibinfo
  {author} {\bibfnamefont {C.}~\bibnamefont {Pfleiderer}}, \bibinfo {author}
  {\bibfnamefont {A.}~\bibnamefont {Rosch}}, \bibinfo {author} {\bibfnamefont
  {A.}~\bibnamefont {Neubauer}}, \bibinfo {author} {\bibfnamefont
  {R.}~\bibnamefont {Georgii}}, \ and\ \bibinfo {author} {\bibfnamefont
  {P.}~\bibnamefont {B{\"o}ni}},\ }\href@noop {} {\bibfield  {journal}
  {\bibinfo  {journal} {Science}\ }\textbf {\bibinfo {volume} {323}},\ \bibinfo
  {pages} {915} (\bibinfo {year} {2009})}\BibitemShut {NoStop}%
\bibitem [{\citenamefont {Neubauer}\ \emph {et~al.}(2009)\citenamefont
  {Neubauer}, \citenamefont {Pfleiderer}, \citenamefont {Binz}, \citenamefont
  {Rosch}, \citenamefont {Ritz}, \citenamefont {Niklowitz},\ and\ \citenamefont
  {B{\"o}ni}}]{neubauer2009}%
  \BibitemOpen
  \bibfield  {author} {\bibinfo {author} {\bibfnamefont {A.}~\bibnamefont
  {Neubauer}}, \bibinfo {author} {\bibfnamefont {C.}~\bibnamefont
  {Pfleiderer}}, \bibinfo {author} {\bibfnamefont {B.}~\bibnamefont {Binz}},
  \bibinfo {author} {\bibfnamefont {A.}~\bibnamefont {Rosch}}, \bibinfo
  {author} {\bibfnamefont {R.}~\bibnamefont {Ritz}}, \bibinfo {author}
  {\bibfnamefont {P.~G.}\ \bibnamefont {Niklowitz}}, \ and\ \bibinfo {author}
  {\bibfnamefont {P.}~\bibnamefont {B{\"o}ni}},\ }\href@noop {} {\bibfield
  {journal} {\bibinfo  {journal} {Physical Review Letters}\ }\textbf {\bibinfo
  {volume} {102}},\ \bibinfo {pages} {186602} (\bibinfo {year}
  {2009})}\BibitemShut {NoStop}%
\bibitem [{\citenamefont {Pfleiderer}\ \emph {et~al.}(2001)\citenamefont
  {Pfleiderer}, \citenamefont {Julian},\ and\ \citenamefont
  {Lonzarich}}]{pfleiderer2001}%
  \BibitemOpen
  \bibfield  {author} {\bibinfo {author} {\bibfnamefont {C.}~\bibnamefont
  {Pfleiderer}}, \bibinfo {author} {\bibfnamefont {S.}~\bibnamefont {Julian}},
  \ and\ \bibinfo {author} {\bibfnamefont {G.~G.}\ \bibnamefont {Lonzarich}},\
  }\href@noop {} {\bibfield  {journal} {\bibinfo  {journal} {Nature}\ }\textbf
  {\bibinfo {volume} {414}},\ \bibinfo {pages} {427} (\bibinfo {year}
  {2001})}\BibitemShut {NoStop}%
\bibitem [{\citenamefont {Doiron-Leyraud}\ \emph {et~al.}(2003)\citenamefont
  {Doiron-Leyraud}, \citenamefont {Walker}, \citenamefont {Taillefer},
  \citenamefont {Steiner}, \citenamefont {Julian},\ and\ \citenamefont
  {Lonzarich}}]{doiron2003}%
  \BibitemOpen
  \bibfield  {author} {\bibinfo {author} {\bibfnamefont {N.}~\bibnamefont
  {Doiron-Leyraud}}, \bibinfo {author} {\bibfnamefont {I.~R.}\ \bibnamefont
  {Walker}}, \bibinfo {author} {\bibfnamefont {L.}~\bibnamefont {Taillefer}},
  \bibinfo {author} {\bibfnamefont {M.~J.}\ \bibnamefont {Steiner}}, \bibinfo
  {author} {\bibfnamefont {S.~R.}\ \bibnamefont {Julian}}, \ and\ \bibinfo
  {author} {\bibfnamefont {G.~G.}\ \bibnamefont {Lonzarich}},\ }\href@noop {}
  {\bibfield  {journal} {\bibinfo  {journal} {Nature}\ }\textbf {\bibinfo
  {volume} {425}},\ \bibinfo {pages} {595} (\bibinfo {year}
  {2003})}\BibitemShut {NoStop}%
\bibitem [{\citenamefont {Pfleiderer}\ \emph {et~al.}(2007)\citenamefont
  {Pfleiderer}, \citenamefont {B{\"o}ni}, \citenamefont {Keller}, \citenamefont
  {R{\"o}{\ss}ler},\ and\ \citenamefont {Rosch}}]{pfleiderer2007}%
  \BibitemOpen
  \bibfield  {author} {\bibinfo {author} {\bibfnamefont {C.}~\bibnamefont
  {Pfleiderer}}, \bibinfo {author} {\bibfnamefont {P.}~\bibnamefont
  {B{\"o}ni}}, \bibinfo {author} {\bibfnamefont {T.}~\bibnamefont {Keller}},
  \bibinfo {author} {\bibfnamefont {U.~K.}\ \bibnamefont {R{\"o}{\ss}ler}}, \
  and\ \bibinfo {author} {\bibfnamefont {A.}~\bibnamefont {Rosch}},\
  }\href@noop {} {\bibfield  {journal} {\bibinfo  {journal} {Science}\ }\textbf
  {\bibinfo {volume} {316}},\ \bibinfo {pages} {1871} (\bibinfo {year}
  {2007})}\BibitemShut {NoStop}%
\bibitem [{\citenamefont {Lee}\ \emph {et~al.}(2009)\citenamefont {Lee},
  \citenamefont {Kang}, \citenamefont {Onose}, \citenamefont {Tokura},\ and\
  \citenamefont {Ong}}]{lee2009}%
  \BibitemOpen
  \bibfield  {author} {\bibinfo {author} {\bibfnamefont {M.}~\bibnamefont
  {Lee}}, \bibinfo {author} {\bibfnamefont {W.}~\bibnamefont {Kang}}, \bibinfo
  {author} {\bibfnamefont {Y.}~\bibnamefont {Onose}}, \bibinfo {author}
  {\bibfnamefont {Y.}~\bibnamefont {Tokura}}, \ and\ \bibinfo {author}
  {\bibfnamefont {N.~P.}\ \bibnamefont {Ong}},\ }\href@noop {} {\bibfield
  {journal} {\bibinfo  {journal} {Physical Review Letters}\ }\textbf {\bibinfo
  {volume} {102}},\ \bibinfo {pages} {186601} (\bibinfo {year}
  {2009})}\BibitemShut {NoStop}%
\bibitem [{\citenamefont {Ritz}\ \emph
  {et~al.}(2013{\natexlab{a}})\citenamefont {Ritz}, \citenamefont {Halder},
  \citenamefont {Wagner}, \citenamefont {Franz}, \citenamefont {Bauer},\ and\
  \citenamefont {Pfleiderer}}]{ritz2013}%
  \BibitemOpen
  \bibfield  {author} {\bibinfo {author} {\bibfnamefont {R.}~\bibnamefont
  {Ritz}}, \bibinfo {author} {\bibfnamefont {M.}~\bibnamefont {Halder}},
  \bibinfo {author} {\bibfnamefont {M.}~\bibnamefont {Wagner}}, \bibinfo
  {author} {\bibfnamefont {C.}~\bibnamefont {Franz}}, \bibinfo {author}
  {\bibfnamefont {A.}~\bibnamefont {Bauer}}, \ and\ \bibinfo {author}
  {\bibfnamefont {C.}~\bibnamefont {Pfleiderer}},\ }\href@noop {} {\bibfield
  {journal} {\bibinfo  {journal} {Nature}\ }\textbf {\bibinfo {volume} {497}},\
  \bibinfo {pages} {231} (\bibinfo {year} {2013}{\natexlab{a}})}\BibitemShut
  {NoStop}%
\bibitem [{\citenamefont {Ritz}\ \emph
  {et~al.}(2013{\natexlab{b}})\citenamefont {Ritz}, \citenamefont {Halder},
  \citenamefont {Franz}, \citenamefont {Bauer}, \citenamefont {Wagner},
  \citenamefont {Bamler}, \citenamefont {Rosch},\ and\ \citenamefont
  {Pfleiderer}}]{ritz2013prb}%
  \BibitemOpen
  \bibfield  {author} {\bibinfo {author} {\bibfnamefont {R.}~\bibnamefont
  {Ritz}}, \bibinfo {author} {\bibfnamefont {M.}~\bibnamefont {Halder}},
  \bibinfo {author} {\bibfnamefont {C.}~\bibnamefont {Franz}}, \bibinfo
  {author} {\bibfnamefont {A.}~\bibnamefont {Bauer}}, \bibinfo {author}
  {\bibfnamefont {M.}~\bibnamefont {Wagner}}, \bibinfo {author} {\bibfnamefont
  {R.}~\bibnamefont {Bamler}}, \bibinfo {author} {\bibfnamefont
  {A.}~\bibnamefont {Rosch}}, \ and\ \bibinfo {author} {\bibfnamefont
  {C.}~\bibnamefont {Pfleiderer}},\ }\href@noop {} {\bibfield  {journal}
  {\bibinfo  {journal} {Physical Review B}\ }\textbf {\bibinfo {volume} {87}},\
  \bibinfo {pages} {134424} (\bibinfo {year} {2013}{\natexlab{b}})}\BibitemShut
  {NoStop}%
\bibitem [{\citenamefont {Pfleiderer}\ \emph {et~al.}(2004)\citenamefont
  {Pfleiderer}, \citenamefont {Reznik}, \citenamefont {Pintschovius},
  \citenamefont {v~Lohneysen} \emph {et~al.}}]{pfleiderer2004}%
  \BibitemOpen
  \bibfield  {author} {\bibinfo {author} {\bibfnamefont {C.}~\bibnamefont
  {Pfleiderer}}, \bibinfo {author} {\bibfnamefont {D.}~\bibnamefont {Reznik}},
  \bibinfo {author} {\bibfnamefont {L.}~\bibnamefont {Pintschovius}}, \bibinfo
  {author} {\bibfnamefont {H.}~\bibnamefont {v~Lohneysen}},  \emph {et~al.},\
  }\href@noop {} {\bibfield  {journal} {\bibinfo  {journal} {Nature}\ }\textbf
  {\bibinfo {volume} {427}},\ \bibinfo {pages} {227} (\bibinfo {year}
  {2004})}\BibitemShut {NoStop}%
\bibitem [{\citenamefont {Pintschovius}\ \emph {et~al.}(2004)\citenamefont
  {Pintschovius}, \citenamefont {Reznik}, \citenamefont {Pfleiderer},\ and\
  \citenamefont {von L{\"o}hneysen}}]{pintschovius2004}%
  \BibitemOpen
  \bibfield  {author} {\bibinfo {author} {\bibfnamefont {L.}~\bibnamefont
  {Pintschovius}}, \bibinfo {author} {\bibfnamefont {D.}~\bibnamefont
  {Reznik}}, \bibinfo {author} {\bibfnamefont {C.}~\bibnamefont {Pfleiderer}},
  \ and\ \bibinfo {author} {\bibfnamefont {H.}~\bibnamefont {von
  L{\"o}hneysen}},\ }\href@noop {} {\bibfield  {journal} {\bibinfo  {journal}
  {Pramana}\ }\textbf {\bibinfo {volume} {63}},\ \bibinfo {pages} {117}
  (\bibinfo {year} {2004})}\BibitemShut {NoStop}%
\bibitem [{\citenamefont {Pfleiderer}\ \emph {et~al.}(1997)\citenamefont
  {Pfleiderer}, \citenamefont {McMullan}, \citenamefont {Julian},\ and\
  \citenamefont {Lonzarich}}]{pfleiderer1997}%
  \BibitemOpen
  \bibfield  {author} {\bibinfo {author} {\bibfnamefont {C.}~\bibnamefont
  {Pfleiderer}}, \bibinfo {author} {\bibfnamefont {G.~J.}\ \bibnamefont
  {McMullan}}, \bibinfo {author} {\bibfnamefont {S.~R.}\ \bibnamefont
  {Julian}}, \ and\ \bibinfo {author} {\bibfnamefont {G.~G.}\ \bibnamefont
  {Lonzarich}},\ }\href@noop {} {\bibfield  {journal} {\bibinfo  {journal}
  {Physical Review B}\ }\textbf {\bibinfo {volume} {55}},\ \bibinfo {pages}
  {8330} (\bibinfo {year} {1997})}\BibitemShut {NoStop}%
\bibitem [{\citenamefont {F{\aa}k}\ \emph {et~al.}(2005)\citenamefont
  {F{\aa}k}, \citenamefont {Sadykov}, \citenamefont {Flouquet},\ and\
  \citenamefont {Lapertot}}]{fak2005}%
  \BibitemOpen
  \bibfield  {author} {\bibinfo {author} {\bibfnamefont {B.}~\bibnamefont
  {F{\aa}k}}, \bibinfo {author} {\bibfnamefont {R.~A.}\ \bibnamefont
  {Sadykov}}, \bibinfo {author} {\bibfnamefont {J.}~\bibnamefont {Flouquet}}, \
  and\ \bibinfo {author} {\bibfnamefont {G.}~\bibnamefont {Lapertot}},\
  }\href@noop {} {\bibfield  {journal} {\bibinfo  {journal} {Journal of
  Physics: Condensed Matter}\ }\textbf {\bibinfo {volume} {17}},\ \bibinfo
  {pages} {1635} (\bibinfo {year} {2005})}\BibitemShut {NoStop}%
\bibitem [{\citenamefont {Yu}\ \emph {et~al.}(2004)\citenamefont {Yu},
  \citenamefont {Zamborszky}, \citenamefont {Thompson}, \citenamefont {Sarrao},
  \citenamefont {Torelli}, \citenamefont {Fisk},\ and\ \citenamefont
  {Brown}}]{yu2004}%
  \BibitemOpen
  \bibfield  {author} {\bibinfo {author} {\bibfnamefont {W.}~\bibnamefont
  {Yu}}, \bibinfo {author} {\bibfnamefont {F.}~\bibnamefont {Zamborszky}},
  \bibinfo {author} {\bibfnamefont {J.~D.}\ \bibnamefont {Thompson}}, \bibinfo
  {author} {\bibfnamefont {J.~L.}\ \bibnamefont {Sarrao}}, \bibinfo {author}
  {\bibfnamefont {M.~E.}\ \bibnamefont {Torelli}}, \bibinfo {author}
  {\bibfnamefont {Z.}~\bibnamefont {Fisk}}, \ and\ \bibinfo {author}
  {\bibfnamefont {S.~E.}\ \bibnamefont {Brown}},\ }\href@noop {} {\bibfield
  {journal} {\bibinfo  {journal} {Physical Review Letters}\ }\textbf {\bibinfo
  {volume} {92}},\ \bibinfo {pages} {086403} (\bibinfo {year}
  {2004})}\BibitemShut {NoStop}%
\bibitem [{\citenamefont {Uemura}\ \emph {et~al.}(2007)\citenamefont {Uemura},
  \citenamefont {Goko}, \citenamefont {Gat-Malureanu}, \citenamefont {Carlo},
  \citenamefont {Russo}, \citenamefont {Savici}, \citenamefont {Aczel},
  \citenamefont {MacDougall}, \citenamefont {Rodriguez}, \citenamefont {Luke}
  \emph {et~al.}}]{uemura2007}%
  \BibitemOpen
  \bibfield  {author} {\bibinfo {author} {\bibfnamefont {Y.~J.}\ \bibnamefont
  {Uemura}}, \bibinfo {author} {\bibfnamefont {T.}~\bibnamefont {Goko}},
  \bibinfo {author} {\bibfnamefont {I.~M.}\ \bibnamefont {Gat-Malureanu}},
  \bibinfo {author} {\bibfnamefont {J.~P.}\ \bibnamefont {Carlo}}, \bibinfo
  {author} {\bibfnamefont {P.~L.}\ \bibnamefont {Russo}}, \bibinfo {author}
  {\bibfnamefont {A.~T.}\ \bibnamefont {Savici}}, \bibinfo {author}
  {\bibfnamefont {A.}~\bibnamefont {Aczel}}, \bibinfo {author} {\bibfnamefont
  {G.~J.}\ \bibnamefont {MacDougall}}, \bibinfo {author} {\bibfnamefont
  {J.~A.}\ \bibnamefont {Rodriguez}}, \bibinfo {author} {\bibfnamefont {G.~M.}\
  \bibnamefont {Luke}},  \emph {et~al.},\ }\href@noop {} {\bibfield  {journal}
  {\bibinfo  {journal} {Nature Physics}\ }\textbf {\bibinfo {volume} {3}},\
  \bibinfo {pages} {29} (\bibinfo {year} {2007})}\BibitemShut {NoStop}%
\bibitem [{\citenamefont {Andreica}\ \emph {et~al.}(2010)\citenamefont
  {Andreica}, \citenamefont {DalmasdeR{\'e}otier}, \citenamefont {Yaouanc},
  \citenamefont {Amato},\ and\ \citenamefont {Lapertot}}]{andreica2010}%
  \BibitemOpen
  \bibfield  {author} {\bibinfo {author} {\bibfnamefont {D.}~\bibnamefont
  {Andreica}}, \bibinfo {author} {\bibfnamefont {P.}~\bibnamefont
  {DalmasdeR{\'e}otier}}, \bibinfo {author} {\bibfnamefont {A.}~\bibnamefont
  {Yaouanc}}, \bibinfo {author} {\bibfnamefont {A.}~\bibnamefont {Amato}}, \
  and\ \bibinfo {author} {\bibfnamefont {G.}~\bibnamefont {Lapertot}},\
  }\href@noop {} {\bibfield  {journal} {\bibinfo  {journal} {Physical Review
  B}\ }\textbf {\bibinfo {volume} {81}},\ \bibinfo {pages} {060412(R)}
  (\bibinfo {year} {2010})}\BibitemShut {NoStop}%
\bibitem [{\citenamefont {Sadykov}\ \emph {et~al.}(2018)\citenamefont
  {Sadykov}, \citenamefont {Pappas}, \citenamefont {Bannenberg}, \citenamefont
  {Dalgliesh}, \citenamefont {Falus}, \citenamefont {Goodway},\ and\
  \citenamefont {Lelievre-Berna}}]{sadykov2018}%
  \BibitemOpen
  \bibfield  {author} {\bibinfo {author} {\bibfnamefont {R.}~\bibnamefont
  {Sadykov}}, \bibinfo {author} {\bibfnamefont {C.}~\bibnamefont {Pappas}},
  \bibinfo {author} {\bibfnamefont {L.~J.}\ \bibnamefont {Bannenberg}},
  \bibinfo {author} {\bibfnamefont {R.~M.}\ \bibnamefont {Dalgliesh}}, \bibinfo
  {author} {\bibfnamefont {P.}~\bibnamefont {Falus}}, \bibinfo {author}
  {\bibfnamefont {C.}~\bibnamefont {Goodway}}, \ and\ \bibinfo {author}
  {\bibfnamefont {E.}~\bibnamefont {Lelievre-Berna}},\ }\href@noop {}
  {\bibfield  {journal} {\bibinfo  {journal} {Journal of Neutron Research}\
  }\textbf {\bibinfo {volume} {20}},\ \bibinfo {pages} {25} (\bibinfo {year}
  {2018})}\BibitemShut {NoStop}%
\bibitem [{SI()}]{SI}%
  \BibitemOpen
  \href@noop {} {\enquote {\bibinfo {title} {{See Supplemental Material for
  more experimental details and additional SANS and NSE results}},}\
  }\BibitemShut {NoStop}%
\bibitem [{\citenamefont {Grigoriev}\ \emph {et~al.}(2005)\citenamefont
  {Grigoriev}, \citenamefont {Maleyev}, \citenamefont {Okorokov}, \citenamefont
  {Chetverikov}, \citenamefont {Georgii}, \citenamefont {B{\"o}ni},
  \citenamefont {Lamago}, \citenamefont {Eckerlebe},\ and\ \citenamefont
  {Pranzas}}]{grigoriev2005}%
  \BibitemOpen
  \bibfield  {author} {\bibinfo {author} {\bibfnamefont {S.~V.}\ \bibnamefont
  {Grigoriev}}, \bibinfo {author} {\bibfnamefont {S.~V.}\ \bibnamefont
  {Maleyev}}, \bibinfo {author} {\bibfnamefont {A.~I.}\ \bibnamefont
  {Okorokov}}, \bibinfo {author} {\bibfnamefont {Y.~O.}\ \bibnamefont
  {Chetverikov}}, \bibinfo {author} {\bibfnamefont {R.}~\bibnamefont
  {Georgii}}, \bibinfo {author} {\bibfnamefont {P.}~\bibnamefont {B{\"o}ni}},
  \bibinfo {author} {\bibfnamefont {D.}~\bibnamefont {Lamago}}, \bibinfo
  {author} {\bibfnamefont {H.}~\bibnamefont {Eckerlebe}}, \ and\ \bibinfo
  {author} {\bibfnamefont {K.}~\bibnamefont {Pranzas}},\ }\href@noop {}
  {\bibfield  {journal} {\bibinfo  {journal} {Physical Review B}\ }\textbf
  {\bibinfo {volume} {72}},\ \bibinfo {pages} {134420} (\bibinfo {year}
  {2005})}\BibitemShut {NoStop}%
\bibitem [{\citenamefont {Pappas}\ \emph {et~al.}(2009)\citenamefont {Pappas},
  \citenamefont {Lelievre-Berna}, \citenamefont {Falus}, \citenamefont
  {Bentley}, \citenamefont {Moskvin}, \citenamefont {Grigoriev}, \citenamefont
  {Fouquet},\ and\ \citenamefont {Farago}}]{pappas2009}%
  \BibitemOpen
  \bibfield  {author} {\bibinfo {author} {\bibfnamefont {C.}~\bibnamefont
  {Pappas}}, \bibinfo {author} {\bibfnamefont {E.}~\bibnamefont
  {Lelievre-Berna}}, \bibinfo {author} {\bibfnamefont {P.}~\bibnamefont
  {Falus}}, \bibinfo {author} {\bibfnamefont {P.~M.}\ \bibnamefont {Bentley}},
  \bibinfo {author} {\bibfnamefont {E.}~\bibnamefont {Moskvin}}, \bibinfo
  {author} {\bibfnamefont {S.}~\bibnamefont {Grigoriev}}, \bibinfo {author}
  {\bibfnamefont {P.}~\bibnamefont {Fouquet}}, \ and\ \bibinfo {author}
  {\bibfnamefont {B.}~\bibnamefont {Farago}},\ }\href@noop {} {\bibfield
  {journal} {\bibinfo  {journal} {Physical Review Letters}\ }\textbf {\bibinfo
  {volume} {102}},\ \bibinfo {pages} {197202} (\bibinfo {year}
  {2009})}\BibitemShut {NoStop}%
\bibitem [{\citenamefont {Janoschek}\ \emph {et~al.}(2013)\citenamefont
  {Janoschek}, \citenamefont {Garst}, \citenamefont {Bauer}, \citenamefont
  {Krautscheid}, \citenamefont {Georgii}, \citenamefont {B{\"o}ni},\ and\
  \citenamefont {Pfleiderer}}]{janoschek2013}%
  \BibitemOpen
  \bibfield  {author} {\bibinfo {author} {\bibfnamefont {M.}~\bibnamefont
  {Janoschek}}, \bibinfo {author} {\bibfnamefont {M.}~\bibnamefont {Garst}},
  \bibinfo {author} {\bibfnamefont {A.}~\bibnamefont {Bauer}}, \bibinfo
  {author} {\bibfnamefont {P.}~\bibnamefont {Krautscheid}}, \bibinfo {author}
  {\bibfnamefont {R.}~\bibnamefont {Georgii}}, \bibinfo {author} {\bibfnamefont
  {P.}~\bibnamefont {B{\"o}ni}}, \ and\ \bibinfo {author} {\bibfnamefont
  {C.}~\bibnamefont {Pfleiderer}},\ }\href@noop {} {\bibfield  {journal}
  {\bibinfo  {journal} {Physical Review B}\ }\textbf {\bibinfo {volume} {87}},\
  \bibinfo {pages} {134407} (\bibinfo {year} {2013})}\BibitemShut {NoStop}%
\bibitem [{\citenamefont {Pappas}\ \emph {et~al.}(2017)\citenamefont {Pappas},
  \citenamefont {Bannenberg}, \citenamefont {Leli{\`e}vre-Berna}, \citenamefont
  {Qian}, \citenamefont {Dewhurst}, \citenamefont {Dalgliesh}, \citenamefont
  {Schlagel}, \citenamefont {Lograsso},\ and\ \citenamefont
  {Falus}}]{pappas2017}%
  \BibitemOpen
  \bibfield  {author} {\bibinfo {author} {\bibfnamefont {C.}~\bibnamefont
  {Pappas}}, \bibinfo {author} {\bibfnamefont {L.~J.}\ \bibnamefont
  {Bannenberg}}, \bibinfo {author} {\bibfnamefont {E.}~\bibnamefont
  {Leli{\`e}vre-Berna}}, \bibinfo {author} {\bibfnamefont {F.}~\bibnamefont
  {Qian}}, \bibinfo {author} {\bibfnamefont {C.~D.}\ \bibnamefont {Dewhurst}},
  \bibinfo {author} {\bibfnamefont {R.~M.}\ \bibnamefont {Dalgliesh}}, \bibinfo
  {author} {\bibfnamefont {D.~L.}\ \bibnamefont {Schlagel}}, \bibinfo {author}
  {\bibfnamefont {T.~A.}\ \bibnamefont {Lograsso}}, \ and\ \bibinfo {author}
  {\bibfnamefont {P.}~\bibnamefont {Falus}},\ }\href@noop {} {\bibfield
  {journal} {\bibinfo  {journal} {Physical Review Letters}\ }\textbf {\bibinfo
  {volume} {119}},\ \bibinfo {pages} {047203} (\bibinfo {year}
  {2017})}\BibitemShut {NoStop}%
\bibitem [{\citenamefont {Bannenberg}\ \emph
  {et~al.}(2017{\natexlab{a}})\citenamefont {Bannenberg}, \citenamefont
  {Kakurai}, \citenamefont {Falus}, \citenamefont {Leli{\`e}vre-Berna},
  \citenamefont {Dalgliesh}, \citenamefont {Dewhurst}, \citenamefont {Qian},
  \citenamefont {Onose}, \citenamefont {Endoh}, \citenamefont {Tokura},\ and\
  \citenamefont {Pappas}}]{bannenberg2017}%
  \BibitemOpen
  \bibfield  {author} {\bibinfo {author} {\bibfnamefont {L.~J.}\ \bibnamefont
  {Bannenberg}}, \bibinfo {author} {\bibfnamefont {K.}~\bibnamefont {Kakurai}},
  \bibinfo {author} {\bibfnamefont {P.}~\bibnamefont {Falus}}, \bibinfo
  {author} {\bibfnamefont {E.}~\bibnamefont {Leli{\`e}vre-Berna}}, \bibinfo
  {author} {\bibfnamefont {R.~M.}\ \bibnamefont {Dalgliesh}}, \bibinfo {author}
  {\bibfnamefont {C.~D.}\ \bibnamefont {Dewhurst}}, \bibinfo {author}
  {\bibfnamefont {F.}~\bibnamefont {Qian}}, \bibinfo {author} {\bibfnamefont
  {Y.}~\bibnamefont {Onose}}, \bibinfo {author} {\bibfnamefont
  {Y.}~\bibnamefont {Endoh}}, \bibinfo {author} {\bibfnamefont
  {Y.}~\bibnamefont {Tokura}}, \ and\ \bibinfo {author} {\bibfnamefont
  {C.}~\bibnamefont {Pappas}},\ }\href@noop {} {\bibfield  {journal} {\bibinfo
  {journal} {Physical Review B}\ }\textbf {\bibinfo {volume} {95}},\ \bibinfo
  {pages} {144433} (\bibinfo {year} {2017}{\natexlab{a}})}\BibitemShut
  {NoStop}%
\bibitem [{\citenamefont {Petrova}\ and\ \citenamefont
  {Stishov}(2012)}]{petrova2012}%
  \BibitemOpen
  \bibfield  {author} {\bibinfo {author} {\bibfnamefont {A.~E.}\ \bibnamefont
  {Petrova}}\ and\ \bibinfo {author} {\bibfnamefont {S.~M.}\ \bibnamefont
  {Stishov}},\ }\href {\doibase 10.1103/PhysRevB.86.174407} {\bibfield
  {journal} {\bibinfo  {journal} {Phys. Rev. B}\ }\textbf {\bibinfo {volume}
  {86}},\ \bibinfo {pages} {174407} (\bibinfo {year} {2012})}\BibitemShut
  {NoStop}%
\bibitem [{\citenamefont {Sidorov}\ \emph {et~al.}(2014)\citenamefont
  {Sidorov}, \citenamefont {Petrova}, \citenamefont {Berdonosov}, \citenamefont
  {Dolgikh},\ and\ \citenamefont {Stishov}}]{sidorov2014}%
  \BibitemOpen
  \bibfield  {author} {\bibinfo {author} {\bibfnamefont {V.~A.}\ \bibnamefont
  {Sidorov}}, \bibinfo {author} {\bibfnamefont {A.~E.}\ \bibnamefont
  {Petrova}}, \bibinfo {author} {\bibfnamefont {P.~S.}\ \bibnamefont
  {Berdonosov}}, \bibinfo {author} {\bibfnamefont {V.~A.}\ \bibnamefont
  {Dolgikh}}, \ and\ \bibinfo {author} {\bibfnamefont {S.~M.}\ \bibnamefont
  {Stishov}},\ }\href@noop {} {\bibfield  {journal} {\bibinfo  {journal}
  {Physical Review B}\ }\textbf {\bibinfo {volume} {89}},\ \bibinfo {pages}
  {100403(R)} (\bibinfo {year} {2014})}\BibitemShut {NoStop}%
\bibitem [{\citenamefont {Bak}\ and\ \citenamefont {Jensen}(1980)}]{bak1980}%
  \BibitemOpen
  \bibfield  {author} {\bibinfo {author} {\bibfnamefont {P.}~\bibnamefont
  {Bak}}\ and\ \bibinfo {author} {\bibfnamefont {M.~H.}\ \bibnamefont
  {Jensen}},\ }\href@noop {} {\bibfield  {journal} {\bibinfo  {journal}
  {Journal of Physics C: Solid State Physics}\ }\textbf {\bibinfo {volume}
  {13}},\ \bibinfo {pages} {L881} (\bibinfo {year} {1980})}\BibitemShut
  {NoStop}%
\bibitem [{\citenamefont {Thessieu}\ \emph {et~al.}(1997)\citenamefont
  {Thessieu}, \citenamefont {Pfleiderer}, \citenamefont {Stepanov},\ and\
  \citenamefont {Flouquet}}]{thessieu1997}%
  \BibitemOpen
  \bibfield  {author} {\bibinfo {author} {\bibfnamefont {C.}~\bibnamefont
  {Thessieu}}, \bibinfo {author} {\bibfnamefont {C.}~\bibnamefont
  {Pfleiderer}}, \bibinfo {author} {\bibfnamefont {A.~N.}\ \bibnamefont
  {Stepanov}}, \ and\ \bibinfo {author} {\bibfnamefont {J.}~\bibnamefont
  {Flouquet}},\ }\href@noop {} {\bibfield  {journal} {\bibinfo  {journal}
  {Journal of Physics: Condensed Matter}\ }\textbf {\bibinfo {volume} {9}},\
  \bibinfo {pages} {6677} (\bibinfo {year} {1997})}\BibitemShut {NoStop}%
\bibitem [{\citenamefont {Petrova}\ \emph {et~al.}(2006)\citenamefont
  {Petrova}, \citenamefont {Krasnorussky}, \citenamefont {Sarrao},\ and\
  \citenamefont {Stishov}}]{petrova2006}%
  \BibitemOpen
  \bibfield  {author} {\bibinfo {author} {\bibfnamefont {A.~E.}\ \bibnamefont
  {Petrova}}, \bibinfo {author} {\bibfnamefont {V.}~\bibnamefont
  {Krasnorussky}}, \bibinfo {author} {\bibfnamefont {J.}~\bibnamefont
  {Sarrao}}, \ and\ \bibinfo {author} {\bibfnamefont {S.~M.}\ \bibnamefont
  {Stishov}},\ }\href@noop {} {\bibfield  {journal} {\bibinfo  {journal}
  {Physical Review B}\ }\textbf {\bibinfo {volume} {73}},\ \bibinfo {pages}
  {052409} (\bibinfo {year} {2006})}\BibitemShut {NoStop}%
\bibitem [{\citenamefont {Otero-Leal}\ \emph {et~al.}(2009)\citenamefont
  {Otero-Leal}, \citenamefont {Rivadulla}, \citenamefont {Saxena},
  \citenamefont {Ahilan},\ and\ \citenamefont {Rivas}}]{otero2009}%
  \BibitemOpen
  \bibfield  {author} {\bibinfo {author} {\bibfnamefont {M.}~\bibnamefont
  {Otero-Leal}}, \bibinfo {author} {\bibfnamefont {F.}~\bibnamefont
  {Rivadulla}}, \bibinfo {author} {\bibfnamefont {S.~S.}\ \bibnamefont
  {Saxena}}, \bibinfo {author} {\bibfnamefont {K.}~\bibnamefont {Ahilan}}, \
  and\ \bibinfo {author} {\bibfnamefont {J.}~\bibnamefont {Rivas}},\
  }\href@noop {} {\bibfield  {journal} {\bibinfo  {journal} {Physical Review
  B}\ }\textbf {\bibinfo {volume} {79}},\ \bibinfo {pages} {060401(R)}
  (\bibinfo {year} {2009})}\BibitemShut {NoStop}%
\bibitem [{\citenamefont {Stishov}\ \emph {et~al.}(2007)\citenamefont
  {Stishov}, \citenamefont {Petrova}, \citenamefont {Khasanov}, \citenamefont
  {Panova}, \citenamefont {Shikov}, \citenamefont {Lashley}, \citenamefont
  {Wu},\ and\ \citenamefont {Lograsso}}]{stishov2007}%
  \BibitemOpen
  \bibfield  {author} {\bibinfo {author} {\bibfnamefont {S.~M.}\ \bibnamefont
  {Stishov}}, \bibinfo {author} {\bibfnamefont {A.~E.}\ \bibnamefont
  {Petrova}}, \bibinfo {author} {\bibfnamefont {S.}~\bibnamefont {Khasanov}},
  \bibinfo {author} {\bibfnamefont {G.~K.}\ \bibnamefont {Panova}}, \bibinfo
  {author} {\bibfnamefont {A.~A.}\ \bibnamefont {Shikov}}, \bibinfo {author}
  {\bibfnamefont {J.~C.}\ \bibnamefont {Lashley}}, \bibinfo {author}
  {\bibfnamefont {D.}~\bibnamefont {Wu}}, \ and\ \bibinfo {author}
  {\bibfnamefont {T.~A.}\ \bibnamefont {Lograsso}},\ }\href@noop {} {\bibfield
  {journal} {\bibinfo  {journal} {Physical Review B}\ }\textbf {\bibinfo
  {volume} {76}},\ \bibinfo {pages} {052405} (\bibinfo {year}
  {2007})}\BibitemShut {NoStop}%
\bibitem [{\citenamefont {Bannenberg}\ \emph
  {et~al.}(2017{\natexlab{b}})\citenamefont {Bannenberg}, \citenamefont {Qian},
  \citenamefont {Dalgliesh}, \citenamefont {Martin}, \citenamefont
  {Chaboussant}, \citenamefont {Schmidt}, \citenamefont {Schlagel},
  \citenamefont {Lograsso}, \citenamefont {Wilhelm},\ and\ \citenamefont
  {Pappas}}]{bannenberg2017reorientations}%
  \BibitemOpen
  \bibfield  {author} {\bibinfo {author} {\bibfnamefont {L.~J.}\ \bibnamefont
  {Bannenberg}}, \bibinfo {author} {\bibfnamefont {F.}~\bibnamefont {Qian}},
  \bibinfo {author} {\bibfnamefont {R.~M.}\ \bibnamefont {Dalgliesh}}, \bibinfo
  {author} {\bibfnamefont {N.}~\bibnamefont {Martin}}, \bibinfo {author}
  {\bibfnamefont {G.}~\bibnamefont {Chaboussant}}, \bibinfo {author}
  {\bibfnamefont {M.}~\bibnamefont {Schmidt}}, \bibinfo {author} {\bibfnamefont
  {D.~L.}\ \bibnamefont {Schlagel}}, \bibinfo {author} {\bibfnamefont {T.~A.}\
  \bibnamefont {Lograsso}}, \bibinfo {author} {\bibfnamefont {H.}~\bibnamefont
  {Wilhelm}}, \ and\ \bibinfo {author} {\bibfnamefont {C.}~\bibnamefont
  {Pappas}},\ }\href@noop {} {\bibfield  {journal} {\bibinfo  {journal}
  {Physical Review B}\ }\textbf {\bibinfo {volume} {96}},\ \bibinfo {pages}
  {184416} (\bibinfo {year} {2017}{\natexlab{b}})}\BibitemShut {NoStop}%
\bibitem [{\citenamefont {R{\"o}{\ss}ler}\ \emph {et~al.}(2006)\citenamefont
  {R{\"o}{\ss}ler}, \citenamefont {Bogdanov},\ and\ \citenamefont
  {Pfleiderer}}]{rossler2006}%
  \BibitemOpen
  \bibfield  {author} {\bibinfo {author} {\bibfnamefont {U.~K.}\ \bibnamefont
  {R{\"o}{\ss}ler}}, \bibinfo {author} {\bibfnamefont {A.~N.}\ \bibnamefont
  {Bogdanov}}, \ and\ \bibinfo {author} {\bibfnamefont {C.}~\bibnamefont
  {Pfleiderer}},\ }\href@noop {} {\bibfield  {journal} {\bibinfo  {journal}
  {Nature}\ }\textbf {\bibinfo {volume} {442}},\ \bibinfo {pages} {797}
  (\bibinfo {year} {2006})}\BibitemShut {NoStop}%
\bibitem [{\citenamefont {Tewari}\ \emph {et~al.}(2006)\citenamefont {Tewari},
  \citenamefont {Belitz},\ and\ \citenamefont
  {Kirkpatrick}}]{BlueQuantumFog2006}%
  \BibitemOpen
  \bibfield  {author} {\bibinfo {author} {\bibfnamefont {S.}~\bibnamefont
  {Tewari}}, \bibinfo {author} {\bibfnamefont {D.}~\bibnamefont {Belitz}}, \
  and\ \bibinfo {author} {\bibfnamefont {T.~R.}\ \bibnamefont {Kirkpatrick}},\
  }\href@noop {} {\bibfield  {journal} {\bibinfo  {journal} {Physical Review
  Letters}\ }\textbf {\bibinfo {volume} {96}},\ \bibinfo {pages} {047207}
  (\bibinfo {year} {2006})}\BibitemShut {NoStop}%
\bibitem [{\citenamefont {Binz}\ \emph {et~al.}(2006)\citenamefont {Binz},
  \citenamefont {Vishwanath},\ and\ \citenamefont {Aji}}]{Binz2006}%
  \BibitemOpen
  \bibfield  {author} {\bibinfo {author} {\bibfnamefont {B.}~\bibnamefont
  {Binz}}, \bibinfo {author} {\bibfnamefont {A.}~\bibnamefont {Vishwanath}}, \
  and\ \bibinfo {author} {\bibfnamefont {V.}~\bibnamefont {Aji}},\ }\href@noop
  {} {\bibfield  {journal} {\bibinfo  {journal} {Physical Review Letters}\
  }\textbf {\bibinfo {volume} {96}},\ \bibinfo {pages} {207202} (\bibinfo
  {year} {2006})}\BibitemShut {NoStop}%
\bibitem [{\citenamefont {Fischer}\ \emph {et~al.}(2008)\citenamefont
  {Fischer}, \citenamefont {Shah},\ and\ \citenamefont {Rosch}}]{Fischer2008}%
  \BibitemOpen
  \bibfield  {author} {\bibinfo {author} {\bibfnamefont {I.}~\bibnamefont
  {Fischer}}, \bibinfo {author} {\bibfnamefont {N.}~\bibnamefont {Shah}}, \
  and\ \bibinfo {author} {\bibfnamefont {A.}~\bibnamefont {Rosch}},\
  }\href@noop {} {\bibfield  {journal} {\bibinfo  {journal} {Physical Review
  B}\ }\textbf {\bibinfo {volume} {77}},\ \bibinfo {pages} {024415} (\bibinfo
  {year} {2008})}\BibitemShut {NoStop}%
\bibitem [{\citenamefont {Hopkinson}\ and\ \citenamefont
  {Kee}(2009)}]{Hopkinson2009}%
  \BibitemOpen
  \bibfield  {author} {\bibinfo {author} {\bibfnamefont {J.~M.}\ \bibnamefont
  {Hopkinson}}\ and\ \bibinfo {author} {\bibfnamefont {H.-Y.}\ \bibnamefont
  {Kee}},\ }\href@noop {} {\bibfield  {journal} {\bibinfo  {journal} {Physical
  Review B}\ }\textbf {\bibinfo {volume} {79}},\ \bibinfo {pages} {14421}
  (\bibinfo {year} {2009})}\BibitemShut {NoStop}%
\bibitem [{\citenamefont {Kirkpatrick}\ and\ \citenamefont
  {Belitz}(2010)}]{Kirkpatrick2010}%
  \BibitemOpen
  \bibfield  {author} {\bibinfo {author} {\bibfnamefont {T.~R.}\ \bibnamefont
  {Kirkpatrick}}\ and\ \bibinfo {author} {\bibfnamefont {D.}~\bibnamefont
  {Belitz}},\ }\href@noop {} {\bibfield  {journal} {\bibinfo  {journal}
  {Physical Review Letters}\ }\textbf {\bibinfo {volume} {104}},\ \bibinfo
  {pages} {256404} (\bibinfo {year} {2010})}\BibitemShut {NoStop}%
\bibitem [{\citenamefont {Pfleiderer}\ \emph {et~al.}(2009)\citenamefont
  {Pfleiderer}, \citenamefont {Neubauer}, \citenamefont {Muhlbauer},
  \citenamefont {Jonietz}, \citenamefont {Janoschek}, \citenamefont {Legl},
  \citenamefont {Ritz}, \citenamefont {M{\"u}nzer}, \citenamefont {Franz},
  \citenamefont {Niklowitz}, \citenamefont {Keller}, \citenamefont {Georgii},
  \citenamefont {B{\"o}ni}, \citenamefont {Binz}, \citenamefont {Rosch},
  \citenamefont {R{\"o}{\ss}ler},\ and\ \citenamefont
  {Bogdanov}}]{Pfleiderer2009dr}%
  \BibitemOpen
  \bibfield  {author} {\bibinfo {author} {\bibfnamefont {C.}~\bibnamefont
  {Pfleiderer}}, \bibinfo {author} {\bibfnamefont {A.}~\bibnamefont
  {Neubauer}}, \bibinfo {author} {\bibfnamefont {S.}~\bibnamefont {Muhlbauer}},
  \bibinfo {author} {\bibfnamefont {F.}~\bibnamefont {Jonietz}}, \bibinfo
  {author} {\bibfnamefont {M.}~\bibnamefont {Janoschek}}, \bibinfo {author}
  {\bibfnamefont {S.}~\bibnamefont {Legl}}, \bibinfo {author} {\bibfnamefont
  {R.}~\bibnamefont {Ritz}}, \bibinfo {author} {\bibfnamefont {W.}~\bibnamefont
  {M{\"u}nzer}}, \bibinfo {author} {\bibfnamefont {C.}~\bibnamefont {Franz}},
  \bibinfo {author} {\bibfnamefont {P.~G.}\ \bibnamefont {Niklowitz}}, \bibinfo
  {author} {\bibfnamefont {T.}~\bibnamefont {Keller}}, \bibinfo {author}
  {\bibfnamefont {R.}~\bibnamefont {Georgii}}, \bibinfo {author} {\bibfnamefont
  {P.}~\bibnamefont {B{\"o}ni}}, \bibinfo {author} {\bibfnamefont
  {B.}~\bibnamefont {Binz}}, \bibinfo {author} {\bibfnamefont {A.}~\bibnamefont
  {Rosch}}, \bibinfo {author} {\bibfnamefont {U.~K.}\ \bibnamefont
  {R{\"o}{\ss}ler}}, \ and\ \bibinfo {author} {\bibfnamefont {A.~N.}\
  \bibnamefont {Bogdanov}},\ }\href@noop {} {\bibfield  {journal} {\bibinfo
  {journal} {Journal of Physics-Condensed Matter}\ }\textbf {\bibinfo {volume}
  {21}},\ \bibinfo {pages} {164215} (\bibinfo {year} {2009})}\BibitemShut
  {NoStop}%
\bibitem [{\citenamefont {Leonov}\ and\ \citenamefont
  {Bogdanov}(2018)}]{leonov2018b}%
  \BibitemOpen
  \bibfield  {author} {\bibinfo {author} {\bibfnamefont {A.~O.}\ \bibnamefont
  {Leonov}}\ and\ \bibinfo {author} {\bibfnamefont {A.~N.}\ \bibnamefont
  {Bogdanov}},\ }\href@noop {} {\bibfield  {journal} {\bibinfo  {journal} {New
  Journal of Physics}\ }\textbf {\bibinfo {volume} {20}},\ \bibinfo {pages}
  {043017} (\bibinfo {year} {2018})}\BibitemShut {NoStop}%
\bibitem [{\citenamefont {Bannenberg}\ \emph
  {et~al.}(2018{\natexlab{a}})\citenamefont {Bannenberg}, \citenamefont
  {Weber}, \citenamefont {Lefering}, \citenamefont {Wolf},\ and\ \citenamefont
  {Pappas}}]{bannenberg2018mnfesisquid}%
  \BibitemOpen
  \bibfield  {author} {\bibinfo {author} {\bibfnamefont {L.~J.}\ \bibnamefont
  {Bannenberg}}, \bibinfo {author} {\bibfnamefont {F.}~\bibnamefont {Weber}},
  \bibinfo {author} {\bibfnamefont {A.~J.~E.}\ \bibnamefont {Lefering}},
  \bibinfo {author} {\bibfnamefont {T.}~\bibnamefont {Wolf}}, \ and\ \bibinfo
  {author} {\bibfnamefont {C.}~\bibnamefont {Pappas}},\ }\href@noop {}
  {\bibfield  {journal} {\bibinfo  {journal} {Physical Review B}\ }\textbf
  {\bibinfo {volume} {98}},\ \bibinfo {pages} {184430} (\bibinfo {year}
  {2018}{\natexlab{a}})}\BibitemShut {NoStop}%
\bibitem [{\citenamefont {Bannenberg}\ \emph
  {et~al.}(2018{\natexlab{b}})\citenamefont {Bannenberg}, \citenamefont
  {Dalgliesh}, \citenamefont {Wolf}, \citenamefont {Weber},\ and\ \citenamefont
  {Pappas}}]{bannenberg2018mnfesisans}%
  \BibitemOpen
  \bibfield  {author} {\bibinfo {author} {\bibfnamefont {L.~J.}\ \bibnamefont
  {Bannenberg}}, \bibinfo {author} {\bibfnamefont {R.~M.}\ \bibnamefont
  {Dalgliesh}}, \bibinfo {author} {\bibfnamefont {T.}~\bibnamefont {Wolf}},
  \bibinfo {author} {\bibfnamefont {F.}~\bibnamefont {Weber}}, \ and\ \bibinfo
  {author} {\bibfnamefont {C.}~\bibnamefont {Pappas}},\ }\href@noop {}
  {\bibfield  {journal} {\bibinfo  {journal} {Physical Review B}\ }\textbf
  {\bibinfo {volume} {98}},\ \bibinfo {pages} {184431} (\bibinfo {year}
  {2018}{\natexlab{b}})}\BibitemShut {NoStop}%
\end{thebibliography}%

\newcommand{\beginsupplement}{%
        \setcounter{table}{0}
        \renewcommand{\thetable}{S\arabic{table}}%
        \setcounter{figure}{0}
        \renewcommand{\thefigure}{S\arabic{figure}}%
        \setcounter{equation}{0}
        \renewcommand{\theequation}{S\arabic{equation}}%
        \setcounter{section}{0}
        \renewcommand{\thesection}{S\arabic{section}}%
       }

\setlength{\abovedisplayskip}{0pt}
\setlength{\belowdisplayskip}{0pt}
\setlength{\headsep}{0pt}
\setlength{\partopsep}{0pt}

\beginsupplement

\clearpage
\maketitle
\normalsize
\section*{Experimental Details}
\subsubsection{Sample and Pressure Cell}
The measurements were performed on a 110~mg single crystal of MnSi of approximately 2$\times$2$\times$3~mm$^3$ in size which originates from the same batch as the samples used in \cite{pappas2017,bannenberg2017reorientations,bannenberg2018mnfesisquid,bannenberg2018mnfesisans}. It was aligned with the [1$\bar{1}$0] crystallographic direction vertical and placed in a cylindrical teflon tube. High pressures were generated using a CuBe/TiZr double-wall clamp cell with a 50 - 50\% mixture of FC770 and FC75 as pressure transmitter. Details of the pressure cell and the modus operandi can be found in ref. \cite{sadykov2018}. 

\subsubsection{SANS}
The SANS measurements were performed at the time-of-flight instrument LARMOR of the ISIS neutron source using neutrons with wavelengths 0.09 $\leq \lambda  \leq$ 1.25~nm. The sample was located 4.4~m from the detector that consists of 80 $^3$He tubes, each 8~mm wide. The magnetic field was applied with a 2T 3D vector field cryomagnet both parallel and perpendicular to the incoming neutron beam. We carefully eliminated the residual magnetic fields by warming up the entire cryomagnet prior to the measurements. All SANS patterns are normalized to standard monitor counts and background corrected using a high temperature measurement. 

\begin{figure*}
\begin{minipage}[t]{0.48\linewidth}
\includegraphics[width=.9\linewidth]{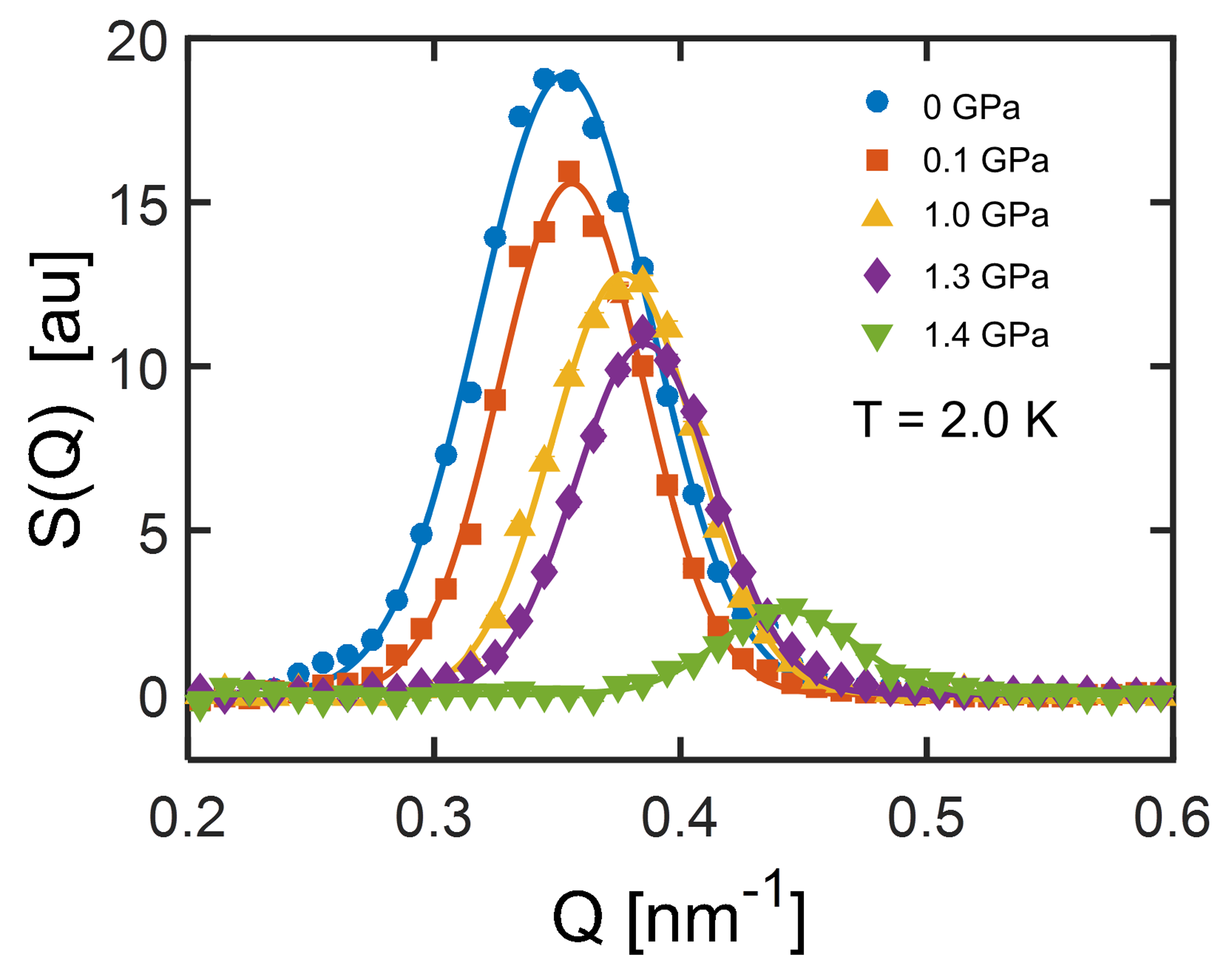}
\caption{Scattering function $S(Q)$ at zero magnetic field, $T$ = 2~K, and for the pressures indicated. $S(Q)$ was obtained by radially averaging the scattered intensity of the 2D SANS patterns. The continuous lines represent fits of Eq. \ref{gauss} to the experimental data.}
\label{SQ}

\end{minipage}\hfill%
\begin{minipage}[t]{0.48\linewidth}
\includegraphics[width= 0.9\textwidth]{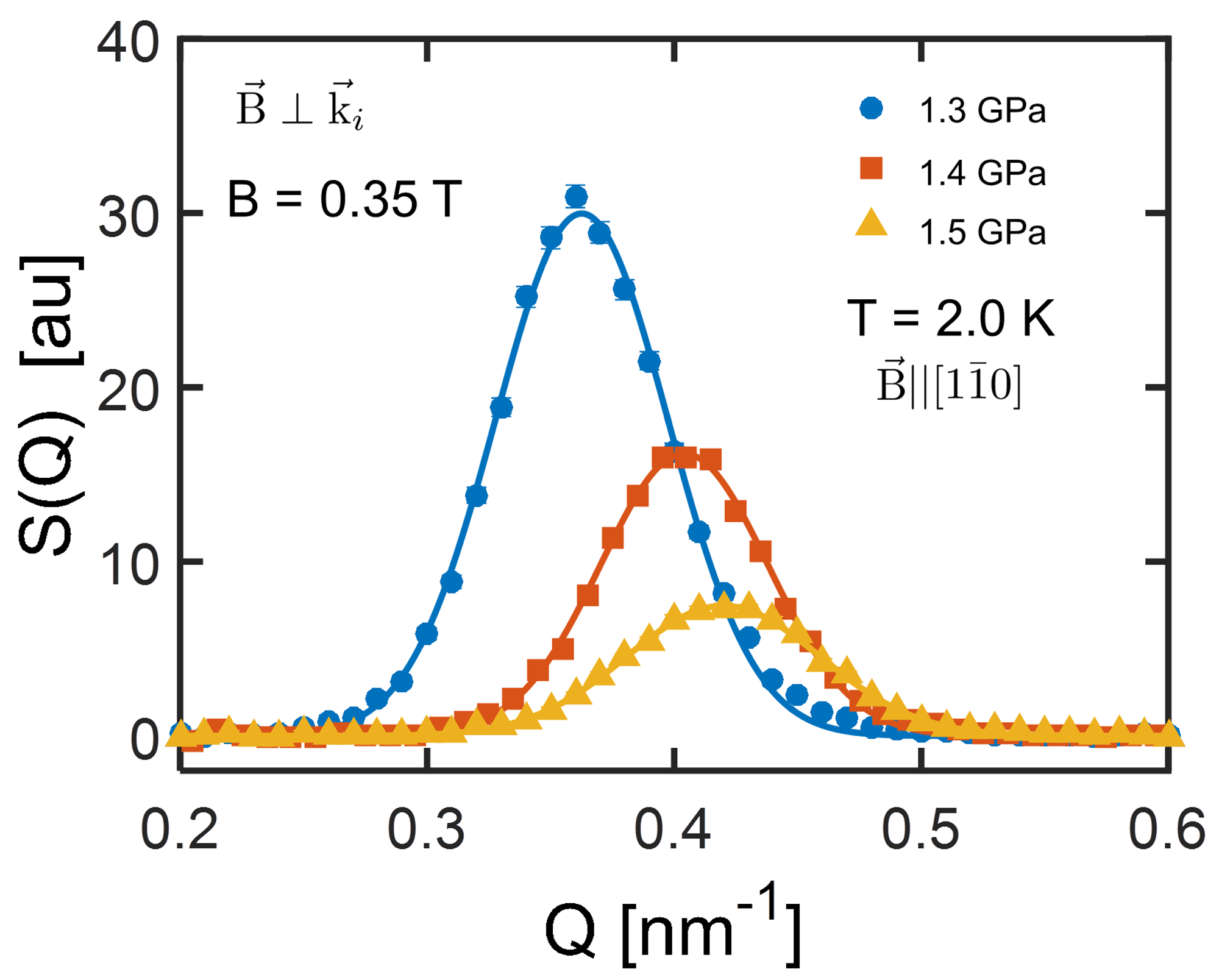}
\caption{Scattering function $S(Q)$ at $B$ = 0.35~T, $T$ = 2~K, for the pressures indicated, and with the magnetic field applied perpendicular to the incoming neutron beam ($\vec{B} \perp \vec{k_i}$). $S(Q)$ was obtained by radially averaging the scattered intensity of the 2D SANS patterns. The continuous lines represent fits of Eq. \ref{gauss} to the experimental data.}
\label{SQ_Bperp_0p35T}
\end{minipage}%
\end{figure*}

The scattering function $S(Q)$ is a one dimensional representation of the 2D SANS patterns and obtained from radially averaging the SANS patterns of e.g. Fig. \ref{ZeroField}. Fig. \ref{SQ} shows example data at zero magnetic field and at $T$ = 2~K for different pressures. The data were fitted to a Gaussian:

\begin{equation}
S(Q) \propto \exp \left(-4 \ln(2) \Big[ (Q-\frac{2\pi}{\ell})/\Delta \Big]^2 \right), 
\label{gauss}
\end{equation}

\noindent with $\ell$ the pitch of the helix and $\Delta$ the FWHM. The resulting values of $\ell$ at zero magnetic field are provided in Fig. \ref{ZeroField}(d). The integrated intensity is computed by summing all scattered intensity with a momentum transfer of 0.2 $<$ $Q$ $<$ 1.2 nm$^{-1}$.

\subsubsection{NSE Spectroscopy}
\begin{figure*}
\begin{minipage}[t]{0.48\linewidth}
\includegraphics[width= 0.9\textwidth]{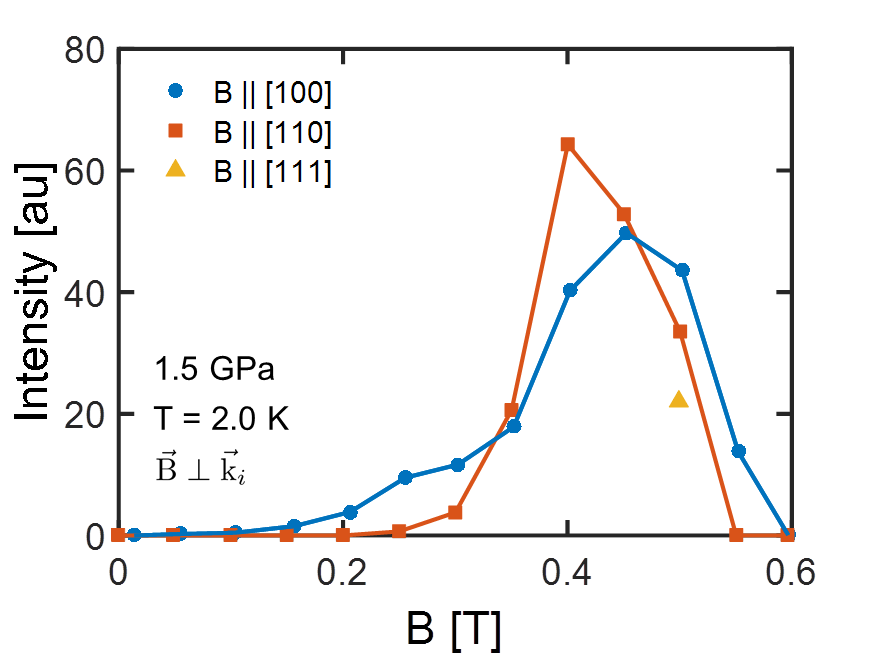}
\caption{Magnetic field dependence of the total amount of magnetic scattering at $T$ = 2~K and $p$ = 1.5~GPa and with the magnetic field applied perpendicular to the incoming neutron beam ($\vec{B} \perp \vec{k_i}$) and applied along the $[001]$, $[1\bar{1}0]$ and $[111]$ crystallographic directions. The measurements were performed by stepwise increasing the magnetic field after ZFC cooling the sample. Unfortunately, only one measurement was performed for $\vec{B} || [111]$ due to beam time constraints.} 
\label{Direction}

\end{minipage}\hfill%
\begin{minipage}[t]{0.48\linewidth}
\includegraphics[width= 0.9\textwidth]{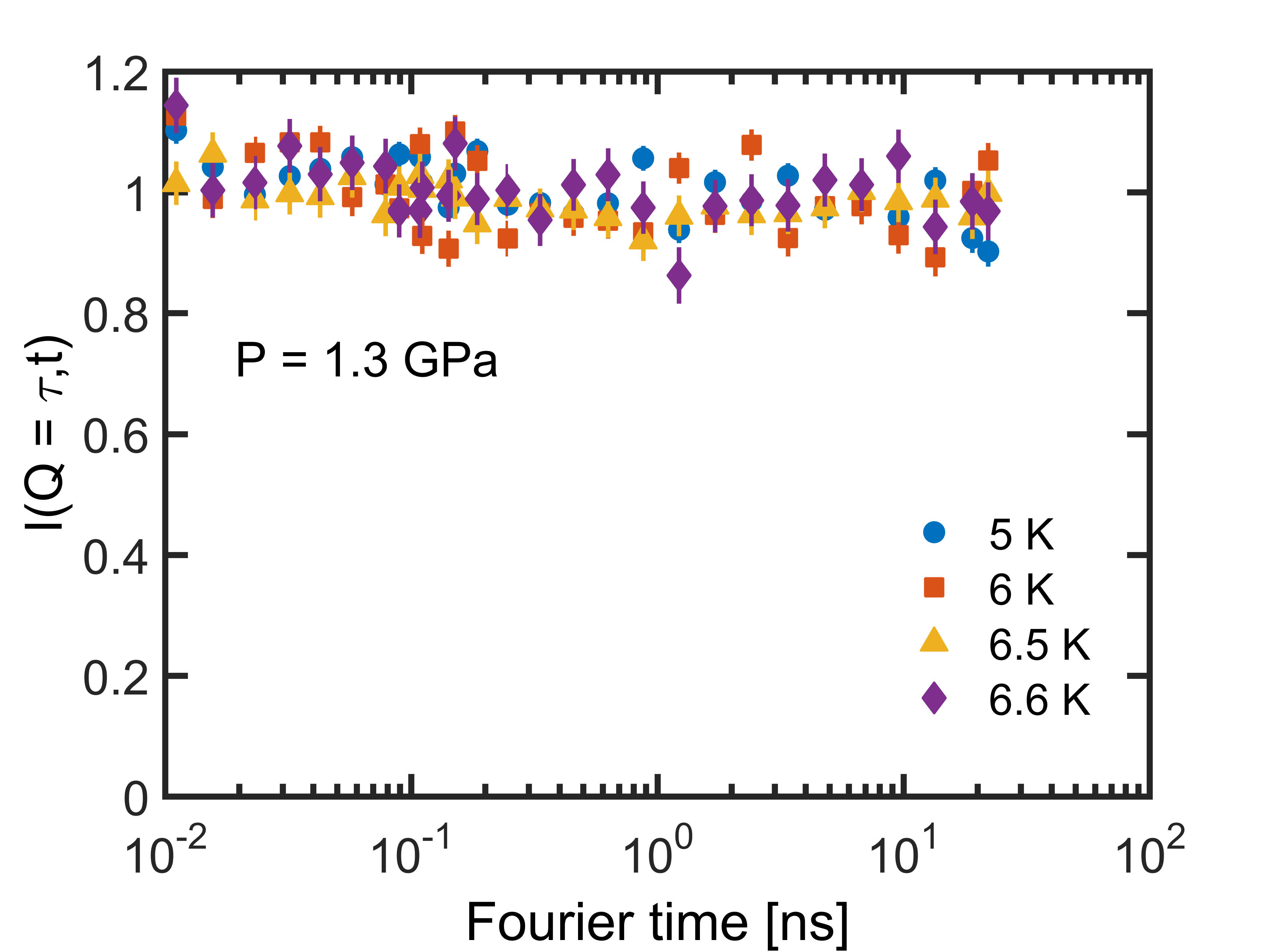}
\caption{Neutron spin echo spectroscopy results obtained at $p$ = 1.3~GPa and at zero magnetic field. The intermediate scattering function $I(Q,t)$ is measured on the magnetic Bragg peak, i.e. $\vec{Q}$ = $\vec{\tau}$, and is provided for four temperatures below the critical temperature of $T_C$ = 6.8~K.}
\label{NSE}
\end{minipage}%
\end{figure*}

The Neutron Spin Echo (NSE) Spectroscopy measurements were performed for $p$ $\sim$ 0.2, 1 and 1.3~GPa at the IN15 spectrometer of the Institute Laue-Langevin. The measurements were performed in the paramagnetic NSE configuration and the neutron beam had a wavelength of  $\lambda$ = 0.9~nm and a monochromatization of $\Delta\lambda$/$\lambda$ = 15\%. The results for $p$=1.3~GPa are displayed in Figure \ref{NSE} and show that the intermediate scattering function $I(Q,t)$ remains purely elastic from low temperatures up to $T$ $\sim$ $T_C$. In fact the NSE signal is elastic at all temperatures where measurements are possible (at higher temperatures the magnetic signal weakens considerably in comparison to the background generated by the pressure cell and this hinders a reliable determination of the paramagnetic NSE response).  

As mentioned in the main text, the absence of relaxation for $T$ $\gtrsim$ $T_C$  confirms the suppression of the precursor phenomena with increasing pressure shown by the SANS results of Fig. \ref{ZeroField}. This suppression is also consistent with the weakening with increasing pressure of the `shoulder', which is characteristic of the precursor phase and is seen slightly above $T_C$ in the temperature dependence of the resistivity and the specific heat \cite{petrova2012,sidorov2014}.

\section*{Supplemental Figures}

\begin{figure*}
\begin{minipage}[t]{0.48\linewidth}
\includegraphics[width=.9\linewidth]{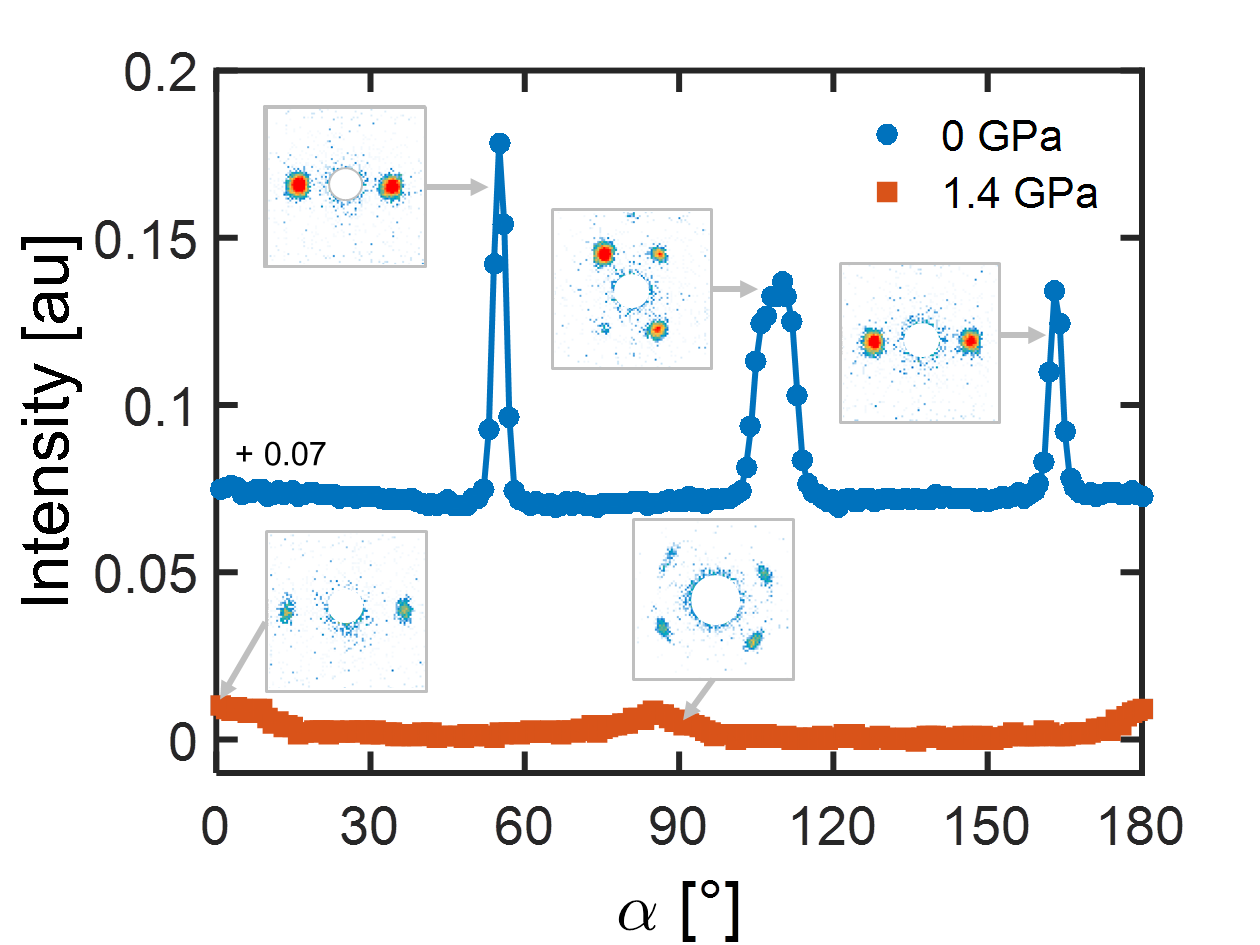}
\caption{ Rocking scan around the vertical $[1\bar{1}0]$ crystallographic direction at $p$ = 0 and 1.4~GPa. $\alpha$ indicates the angle between the $[001]$ crystallographic direction and the SANS detector plane. The data at $p$ = 0.0~GPa are shifted by the value indicated in the plot.}
\label{RockingCurve}

\end{minipage}\hfill%
\begin{minipage}[t]{0.48\linewidth}
\end{minipage}%
\end{figure*}

\begin{figure*}[tb]
\begin{center}
\includegraphics[width= 1\textwidth]{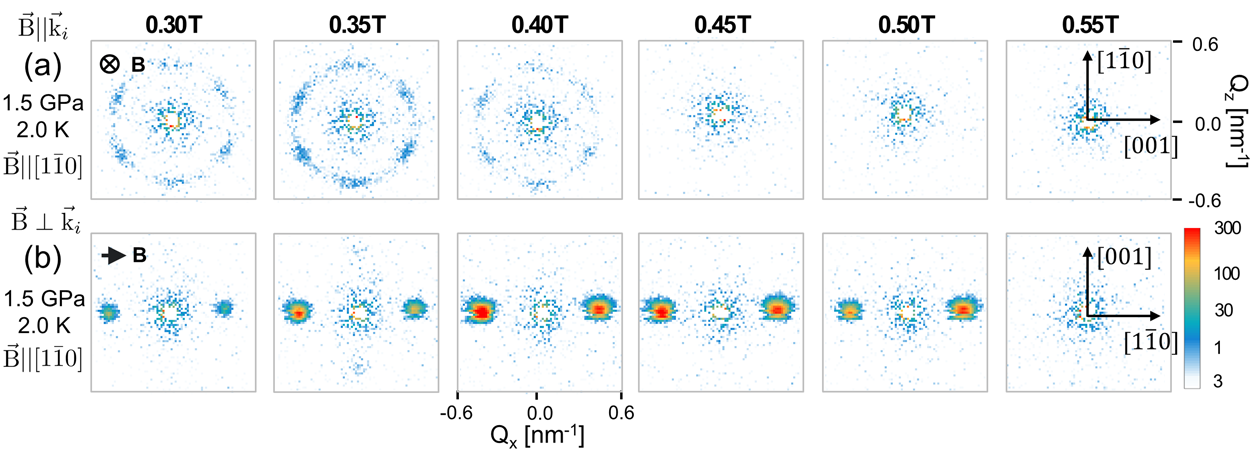}
\caption{Characteristic SANS patterns at $p$ = 1.5~GPa, $T$ = 2~K, and for the magnetic fields indicated. The magnetic field was applied along the $[1\bar{1}0]$ crystallographic direction and both parallel ($\vec{B} || \vec{k_i}$) and perpendicular ($\vec{B} \perp \vec{k_i}$) to the incoming neutron beam.}
\label{Patterns_2K_1p5Gpa}
\end{center}
\end{figure*}

\begin{figure*}
\begin{minipage}[t]{0.48\linewidth}
\includegraphics[width= 0.9\textwidth]{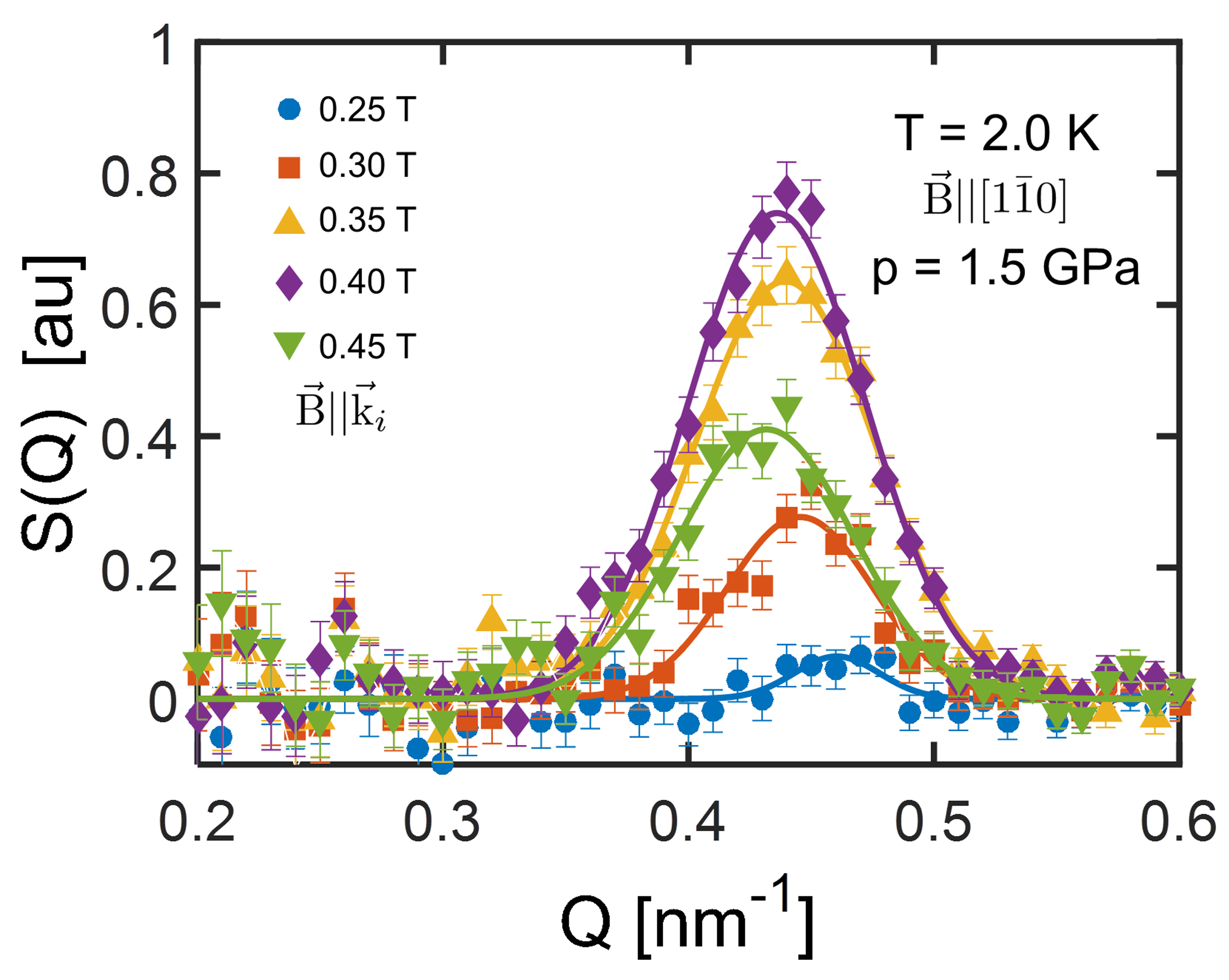}
\caption{Scattering function $S(Q)$ at $p$ = 1.5~GPa, $T$ = 2~K, and for the magnetic fields indicated. The magnetic field was applied along the $[1\bar{1}0]$ crystallographic direction and parallel  to the incoming neutron beam ($\vec{B} || \vec{k_i}$). $S(Q)$ was obtained by radially averaging the scattered intensity of the 2D SANS patterns. The continuous lines represent fits of Eq. \ref{gauss} to the Experimental data.}
\label{SQ_Bperp_0p35T_par}

\end{minipage}\hfill%
\begin{minipage}[t]{0.48\linewidth}
\includegraphics[width= 0.9\textwidth]{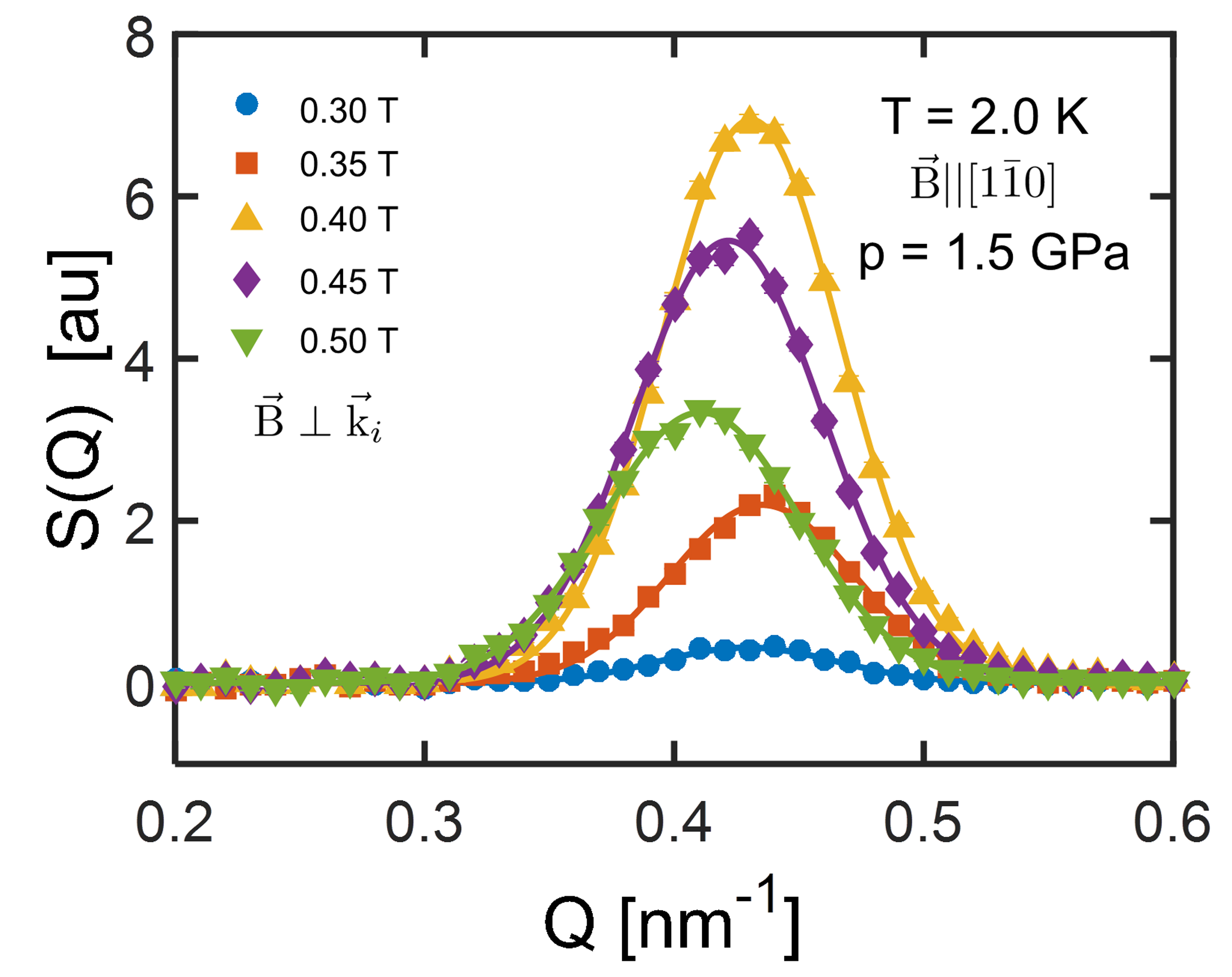}
\caption{Scattering function $S(Q)$ at $p$ = 1.5~GPa, $T$ = 2~K, and for the magnetic fields indicated. The magnetic field was applied along the $[1\bar{1}0]$ crystallographic direction and perpendicular  to the incoming neutron beam ($\vec{B} \perp \vec{k_i}$). $S(Q)$ was obtained by radially averaging the scattered intensity of the 2D SANS patterns. The continuous lines represent fits of Eq. \ref{gauss} to the Experimental data.}
\label{SQ_2K_1p5Gpa}
\end{minipage}%
\end{figure*}

\begin{figure*}
\begin{minipage}[t]{0.48\linewidth}
\includegraphics[width= 0.9\textwidth]{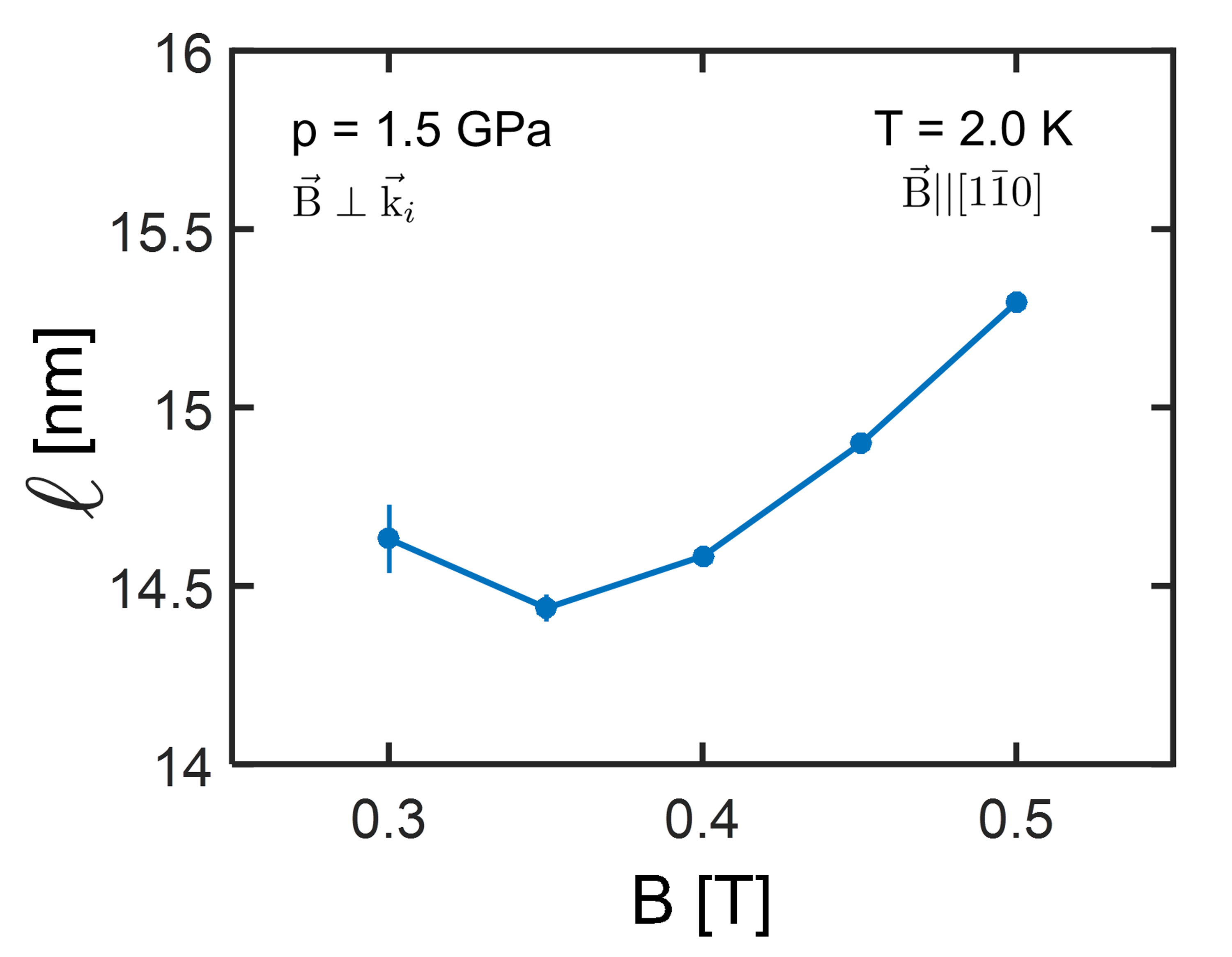}
\caption{Magnetic field dependence of the helimagnetic pitch $\ell$ at $p$ = 1.5~GPa and $T$ = 2~K. The magnetic field was applied along the $[1\bar{1}0]$ crystallographic direction and perpendicular  to the incoming neutron beam ($\vec{B} \perp \vec{k_i}$). The pitch was determined from a fit of equation Eq. \ref{gauss} to $S(Q)$, as displayed in Fig. \ref{Pitch_2K_1p5Gpa}. }
\label{Pitch_2K_1p5Gpa}

\end{minipage}\hfill%
\begin{minipage}[t]{0.48\linewidth}
\includegraphics[width= 0.9\textwidth]{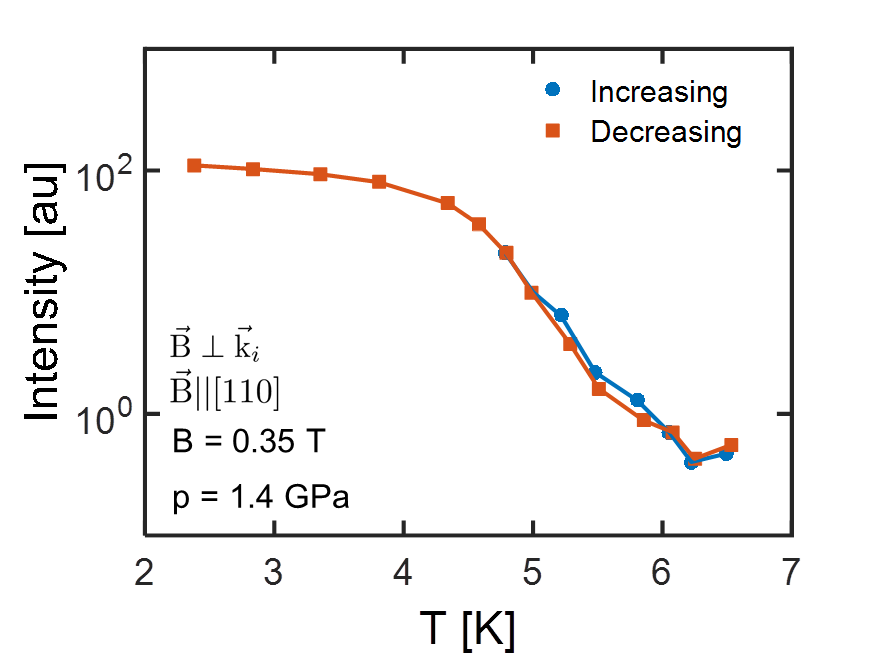}
\caption{Temperature dependence of the total amount of magnetic scattering at $B$ = 0.35~T and $p$ = 1.4~GPa and with the magnetic field applied perpendicular to the incoming neutron beam ($\vec{B} \perp \vec{k_i}$) and applied along the $[1\bar{1}0]$ crystallographic direction. The measurements were performed by stepwise increasing and decreasing the temperature.} 
\label{Temp_History}
\end{minipage}%
\end{figure*}

\end{document}